\documentclass[aps,prd,preprint,superscriptaddress,amsmath,amssymb,showpacs]{revtex4-1}
\usepackage{dcolumn}
\usepackage{graphicx}
\usepackage{float}
\usepackage{physics}
\usepackage[colorlinks=true,allcolors=blue]{hyperref}
\usepackage{epstopdf}
\usepackage{changes}
\usepackage{xcolor}
\usepackage[T1]{fontenc}
\usepackage{verbatim}

\begin{document}

\title{ Doubly-heavy tetraquark at finite temperature in a holographic model}
\author{Xi Guo }
\affiliation{School of Nuclear Science and Technology, University of South China, Hengyang 421001, China}

\author{Jia-Jie Jiang}
\affiliation{School of Nuclear Science and Technology, University of South China, Hengyang 421001, China}

\author{Xuan Liu}
\affiliation{School of Nuclear Science and Technology, University of South China, Hengyang 421001, China}

\author{Xun Chen}
\email{chenxunhep@qq.com}
\affiliation{School of Nuclear Science and Technology, University of South China, Hengyang 421001, China}

\author{Dong Xiang}
\email{xiangdong@usc.edu.cn}
\affiliation{School of Nuclear Science and Technology, University of South China, Hengyang 421001, China}

\author{Xiao-Hua Li}
\email{lixiaohuaphysics@126.com}
\affiliation{School of Nuclear Science and Technology, University of South China, Hengyang 421001, China}

\date{\today}

\begin{abstract}
In this paper, we employ gauge/gravity duality to investigate the string breaking and melting of doubly-heavy tetraquark that includes two heavy quarks and two light antiquarks in a holographic model at finite temperature. Firstly, four different configurations of $\rm{QQ\bar{q}\bar{q}}$ are studied at different separation distances of the heavy quarks at finite temperatures. At high temperature, $\rm{QQ\bar{q}\bar{q}}$ will melt at certain distances and the screening distance has been given for different temperatures. As the temperature continues to increase, some configurations of doubly-heavy tetraquark can not exist. Furthermore, we investigate three decay modes of $\rm{QQ\bar{q}\bar{q}}$ and compare the potential energy of $\rm{QQ\bar{q}\bar{q}}$ with that of $\rm{QQq}$ at finite temperature.
\end{abstract}

\maketitle
\section{Introduction}\label{sec:01_intro}
The gauge/gravity duality has been widely recognized as a fundamental feature of quantum gravity, and extensive research has been carried out in this field over the past two decades, leading to many important findings \cite{1Maldacena:1998TheLargeN}. Holographic QCD offers a novel approach to study and compute the properties of various physical phenomena in QCD. Studying heavy quarkonium is beneficial for understanding the properties of quark-gluon plasma($\rm{QGP}$), as well as validate our understanding of the interactions between hadrons and fundamental particles \cite{2Aharony:1999LargeN,3Maldacena:1998}. In holographic QCD, we can study the interaction between quarks and antiquarks by placing a pair of them inside a bulk. Then, we can utilize the tools of string theory to calculate the interaction between the two quarks and their corresponding potential energy.

The string breaking phenomenon is a result of the nonperturbative effects of QCD. Up to now, lattice QCD has been effectively utilized to investigate this phenomenon, albeit limited to meson modes at zero temperature and zero chemical potential\cite{14Andreev:2019cbc}. As is widely acknowledged, the environment in which tetraquark are situated can often be complex. Therefore, introducing high-temperature factors into the model to study the behavior of four quark potentials at elevated temperatures can aid in better understanding the behavior of tetraquark under extreme conditions. In previous studies, it has been discovered that under sufficiently high temperatures, thermal excitations produce a plasma of quarks and gluons\cite{4Rey:1998Wilson-Polyakovloopwendu}. Subsequently, they discussed the potential at finite temperatures
\cite{4Rey:1998Wilson-Polyakovloopwendu,5Brandhuber:1998limitwendu,6Casalderrey-2014wendu}.  The deformed $\rm{AdS_{5}}$ model \cite{Zhang:2015faa,Andreev:2006ct,Fadafan:2011gm} and Einstein-Maxwell-Dilaton model are employed to compute the quark-antiquark potential \cite{DeWolfe:2010he,Chen:2021gop,Zhou:2020ssi,Chen:2019rez,Bohra:2019ebj,Alho:2013hsa,Chen:2020ath,Erlich:2005wuwei,Chen:2018vty,Arefeva:2018hyo,Ewerz:2016zsx,Casalderrey-Solana:2011dxg,Fang:2015ytf,Ding:2015ona,Zhou:2021sdy,Cai:2012xh,Li:2011hp,Li:2012ay} in these researches.

Recently, at the Large Hadron Collider, the LHCb collaboration observed a hadron state that contains four quarks \cite{Richard:2018jkw,7LHCb:2021vvq}. This tetraquark contains two charm quarks, a $\bar{u}$ and a $\bar{d}$ quark, with a mass of about $3875 $  $\rm{MeV/c^{2}}$. This finding has renewed interest in studying the tetraquark theory. It is worth noting that lattice gauge theory is still a fundamental tool for studying non-perturbative phenomena in QCD, but research results on the potential of $\rm{QQ\bar{q}\bar{q}}$ are limited \cite{Drummond:1998ar,8tetraquark2017,9Andreev:2008some,10Najjar:2009da,11Yamamoto:2008jz,12Francis:2017bjr}.

The tetraquark model used in this paper was proposed by Andreev \cite{13Andreev4kuake}. This model assumes that the heavy quarks are significantly heavier than the typical energy scale of the system, allowing us to treat them as static.  The interaction potential between the quarks is determined by their relative separation. The main reason for studying this model is that its results on both quark-antiquark and tetraquark potentials are consistent with lattice calculations and QCD phenomenology \cite{14Andreev:2019cbc,15Andreev:2015iaa}. At zero temperature, the timelike Wilson loop is realized by a U-shaped macroscopic string for any interquark separation. At finite temperature, we have seen that the string configuration is either a pair of straight strings or U-shaped\cite{4Rey:1998Wilson-Polyakovloopwendu}. Our technique for extracting the potential of $\rm{QQ\bar{q}\bar{q}}$ is similar to the one used in lattice QCD. We extract the potential from the expectation value of the $\rm{QQ\bar{q}\bar{q}}$ Wilson loop, $\rm{W_{QQ\bar{q}\bar{q}}}$$(R, T)$. The $\rm{QQ\bar{q}\bar{q}}$ Wilson loop consists of heavy quark paths and light quark propagators. The separation distance between heavy quark pairs and the potential energy relationship of the model were calculated and analyzed at finite temperature \cite{Stringbreaking,LiYaodong2023,YiYang,Chen:Movingheavyquarkonium,Chen:2019Quarkyonicphase,Zhou:2020Thermodynamicsof,Andreev:2006nw,Andreev:2006TheSpatialString,HeSong:2010Heavyquarkpotential}. Subsequently, three decay modes of tetraquark were studied to determine the most probable decay mode that occurs at high temperatures and to compare it with the decay of three quarks.

Sec.~\ref{sec:02}  provides a brief review of the theoretical foundation of the model and establishes a framework for studying $\rm{QQ\bar{q}\bar{q}}$ at finite temperature. Sec.~\ref{sec:03}  involves numerical solutions for the energy and separation distance between heavy quarks for these configurations at different temperatures. This is followed by a discussion of the results  in Sec.~\ref{sec:04}. In this section, we will also discuss the melting of $\rm{QQ\bar{q}\bar{q}}$ strings at finite temperatures by analyzing potential energy trends. Later in Sec.~\ref{sec:05}, the three types of decay modes in the $\rm{QQ\bar{q}\bar{q}}$ model will be discussed and compared with the decay modes in the QQq model.  Finally, Sec.~\ref{sec:06} presents the summary and conclusion of this paper.

\section{PRELIMINARIES}\label{sec:02}

In this paper, we extend the study of the potential for doubly-heavy tetraquarks from zero temperature \cite{13Andreev4kuake} to finite temperatures. To begin our discussion on the potential of $\rm{QQ\bar{q}\bar{q}}$ at finite temperatures, let us first review the specific holographic model utilized in this paper. Following
 \cite{13Andreev4kuake,Andreev3kuake,Chen3kuake}, the background metric at finite temperatures is given by:
\begin{gather}
d \mathbf{s}^2=\mathrm{e}^{s r^2} \frac{R^2}{r^2}\left(f(r) d t^2+d \vec{x}^2+f^{-1}(r) d r^2\right)+\mathrm{e}^{-s r^2} g_{a b}^{(5)} d \omega^a d \omega^b,
\end{gather}
such model is a deformation of the Euclidean  $\rm{AdS_{5}}$ space of radius $R$, with a deformation parameter  $s$ \cite{Andreev:2007zv}. In the five-dimensional compact space (sphere) characterized by the blackening factor $f$ with coordinates $\omega^a$ and $f(r)$, the Nambu-Goto action of a string is expressed as
\begin{gather}
S_{NG}=\frac{1}{2 \pi \alpha^{\prime}} \int_0^1 d \sigma \int_0^T d \tau \sqrt{\gamma},
\end{gather}
here, $\gamma $ represents an induced metric on the string world-sheet with a Euclidean signature, while $\alpha ^{'}$ is a parameter associated with the string.  For the $\rm{AdS_{5}}$ space, we assume that the blackening factor $f$ takes the form $f(r)=1-\left(\frac{r}{r_h}\right)^4$, where $f(0)=1$ at the boundary and $f\left(r_h\right)=0$ at the horizon. The Hawking temperature, which is consistent with the temperature of the dual gauge theory, can be defined as $T=\frac{1}{4 \pi}\left|\partial_r f\right|_{r=r_h}$.

Then, we consider baryon vertices which are string junctions \cite{Witten:1998xy}. According to the AdS/CFT correspondence, they are represented by a five-brane wrapped around the internal space X at the point where three strings intersect and appear to be joined together in five dimensions \cite{Witten:1998zhongzidingdian,Aharony:1998fanzhongzidingdian}. Correspondingly, the antibaryon vertex is represented by an antibrane in the AdS/CFT correspondence. At leading order $\alpha ^{'}$, the brane's action is $S_{\text {vert }}=\mathcal{T}_5 \int d^6 \xi \sqrt{\gamma^{(6)}}$, where $ \mathcal{T}_5 $ represents the brane tension and $\xi^i$ denotes the world-volume coordinates. If we choose static specifications $\xi^0=t$ and $\xi^a=\theta _{a}$, where $\theta _{a}$ represents the coordinates on X, then the resulting action is as follows
\begin{gather}
S_{\mathrm{vert}}=\tau_v \int d t \frac{\mathrm{e}^{-2 s r^2}}{r} \sqrt{f(r)}.
\end{gather}
Here, $\mathcal{T}_v$ is a dimensionless parameter defined by $\mathcal{T}_v =\mathcal{T}_5Rvol(\rm{X})$, where $vol(\rm{X})$ represents the volume of X, and it serves as a free parameter. Additionally, we have the same action for both baryon and antibaryon vertices $S_{\bar{v}}$ = $S_{v}$ at finite temperature.

Finally, we consider the light quark located at the end of the string,  a scalar field that is coupled to the boundary of the worldsheet via the open string tachyon $S_{q} =\int \mathrm{d} \tau e \rm{T}$ \cite{Andreev:2020pqy}. Here, the integral is over a world-sheet boundary parameterized by $\tau$ ,and $e$ is a boundary metric. Assuming a constant background given by $\rm{T(x,r)=T_{0}}$ and worldsheets with boundaries along lines in the $t$ direction. Thus, the action can be written as
\begin{gather}
S_{\mathrm{q}}=\mathrm{m} \int d t \frac{\mathrm{e}^{\frac{s}{2} r^2}}{r} \sqrt{f(r)}.
\end{gather}
Here, $\rm{m}=$$R$$\rm{T_{0}}$ represents the mass of a point particle at rest, with $\rm{T_{0}}$ as its mass. Therefore, the given action describes the behavior of a point particle with mass $\rm{T_{0}}$ at rest. It should be noted that the same action also describes the behavior of light antiquarks located at string endpoints, and hence $S_{\bar q}$ = $S_{ q}$. In this model, the interaction of quarks is described by the string tension, which is consistent with Ref. \cite{3Maldacena:1998}. 

The model parameters are selected as follows: $g=\frac{R^{2} }{2\pi \alpha ^{'} }$, $\mathrm{k}=\frac{\tau_v}{3\mathrm{~g}}$  and  $n=\frac{m}{g}$.   All our parameters are consistent with Refs.\cite{14Andreev:2019cbc,13Andreev4kuake,Andreev3kuake,Chen3kuake} based on the lattice QCD. The value of $s$ is fixed from the slope of the Regge trajectory of $\rho(n)$ mesons in the soft wall model with the geometry Eq. (1). This gives $s = 0.45 \rm{GeV^2}$. Then, fitting the value of the string tension $\sigma$ to its value in \cite{Stringbreaking} gives $g = 0.176$. According to the gauge/string duality $g$ is related to the 't Hooft coupling. Next, the parameter $m$ is adjusted to reproduce the lattice result for the $\rm{\bar{Q}Q}$ string breaking distance $L_c^{(m)}$. With $L_c^{(m)}=1.22 \mathrm{fm}$\cite{Stringbreaking}, it gives $m=0.538$. The parameter $n$ is expressed in terms of the parameters of as $\mathrm{n}=\frac{\mathrm{m}}{\mathrm{g}}$\cite{14Andreev:2019cbc}. For fixed the value of $k$, one should keep in mind two things. First, the value of $k$ can be adjusted to fit the lattice data for the three-quark potential, as is done in \cite{15Andreev:2015iaa} for pure $S U(3)$ gauge theory. Unfortunately, at the moment, there are no lattice data available for QCD with two light quarks. Second, the range of allowed values for $k$ is limited to $-\frac{\mathrm{e}^3}{15} $ to $-\frac{1}{4} \mathrm{e}^{\frac{1}{4}}$. We take $k=-\frac{1}{4} \mathrm{e}^{\frac{1}{4}}$ simply because it yields an exact solution to equations (20). Finally, the parameters of this article are as follows: $s=0.45$ $\rm{{GeV^2}}$, $g= 0.176$,  $n = 3.057$,  $k =-\frac{1}{4} E^{\frac{1}{4}}$, and  $c=0.623$ $\rm{GeV}$ in this model from Ref. \cite{13Andreev4kuake} at zero temperature. No other extra parameters are introduced in this article.

\section{the connected string configuration of  \texorpdfstring {$\rm{QQ\bar{q}\bar{q}}$}{} }\label{sec:03}

\subsection{Small L}

\begin{figure}
    \centering
    \includegraphics[width=10cm]{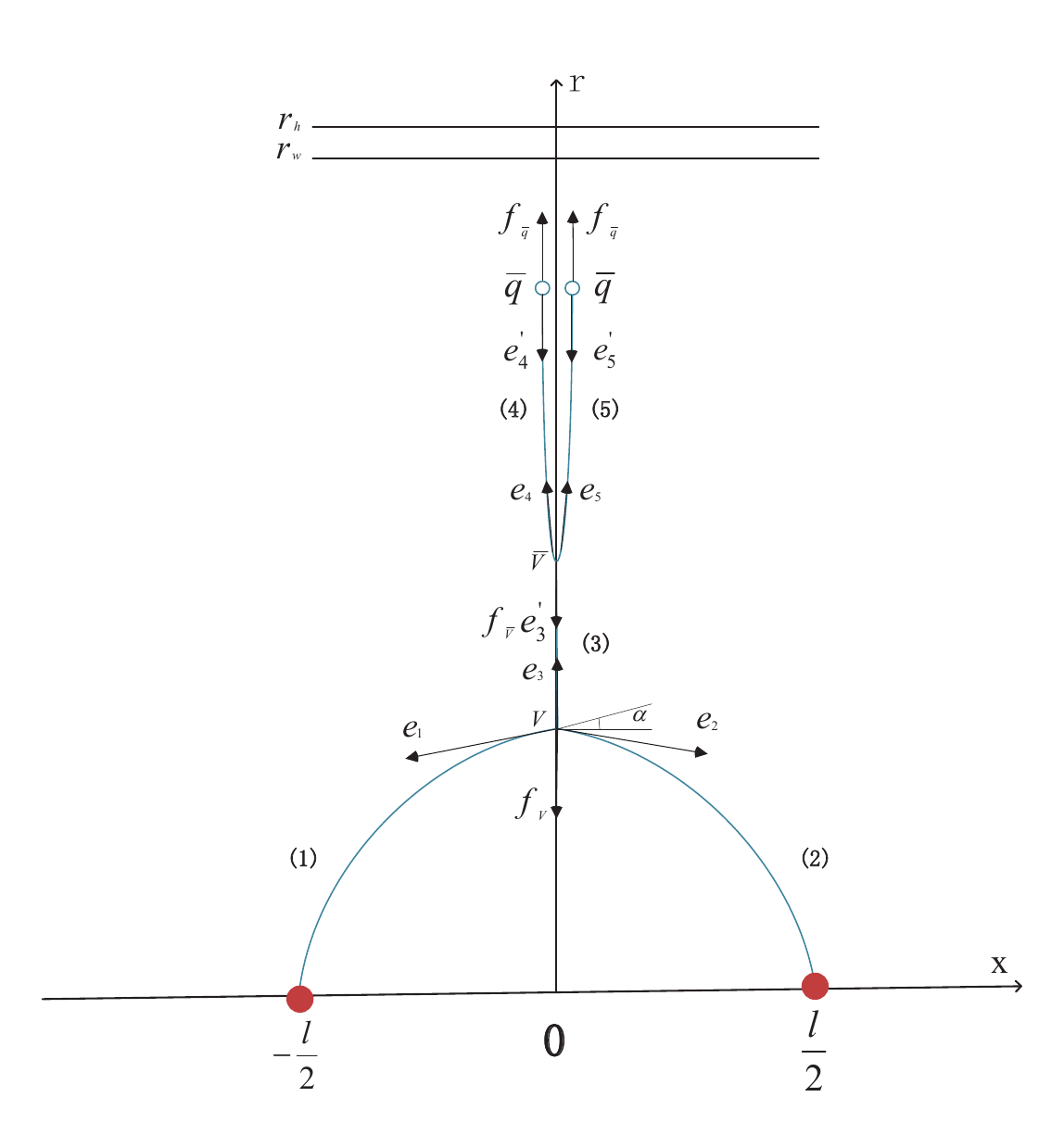}
    \caption{\label{fig1}  A static string configuration at small heavy-quark separation distance. The heavy quarks Q are located on the boundary, whereas the light antiquarks $\bar{q}$, baryon vertex $\rm{V}$, and antibaryon vertex $\rm{\bar{V}}$ are situated in the bulk of the five-dimensional space at $r_{\bar{q}}$, $r_v$ and $r_{\bar{v}}$, respectively. $r_h$ represents the position of the black-hole horizon, while $r_w$ indicates the position of a dynamic wall in the confined phase.}
\end{figure}

The configuration of $\rm{QQ\bar{q}\bar{q}}$ is illustrated in Fig.\ref{fig1}. The total action for this system is given by the sum of the Nambu-Goto actions, as well as the actions associated with the vertices and antiquarks.
\begin{gather}
 S=\sum_{i=1}^5 S_{\mathrm{NG}}^{(i)}+2 S_{\mathrm{v}}+2 S_{\mathrm{\bar{q}}}.
\end{gather}
We choose the static gauge $\xi^1=t$ and  $\xi^2=r$ for the Nambu-Goto actions, the boundary conditions for $x(r)$ become

\begin{gather}
 x^{(1)}(0)=-\frac{1}{2} \ell, \quad x^{(2)}(0)=\frac{1}{2} \ell, \quad
x^{(1,2,3)}\left(r_v\right)=x^{(3,4,5)}\left(r_{\bar{v}}\right)=x^{(4,5)}\left(r_{\bar{q}}\right)=0.
\end{gather}
Considering the boundary conditions, we get the total action
\begin{gather}
\begin{aligned}
S & =g T\left(2 \int_0^{r_v} \frac{e^{s r^2}}{r^2} \sqrt{1+f(r)\left(\partial_r x\right)^2} \mathrm{~d} r+\int_{r_v}^{r_{\bar{v}}} \frac{e^{s r^2}}{r^2} \mathrm{~d} r+2 \int_{r_{\bar{v}}}^{r_{\bar{q}}} \frac{e^{s r^2}}{r^2} \mathrm{~d} r\right. \\
& +3 k \frac{e^{-2 s r_{\bar{v}}^2}}{r_{\bar{v}}} \sqrt{f\left(r_{\bar{v}}\right)}+3 k \frac{e^{-2 s r_v^2}}{r_v} \sqrt{f\left(r_v\right)}+2 n \frac{e^{\frac{1}{2} r_{\bar{q}}^2}}{r_{\bar{q}}} \sqrt{\left.f\left(r_{\bar{q}}\right)\right)} .
\end{aligned}
\end{gather}
Here $ \partial _{r} x=\frac{\partial x }{\partial r}$ , $T=\int_{0}^{T} \mathrm{d}t $ and the straight strings are located at $x=0$. For string (1) and (2), which correspond to the first term in (7), we can derive the equation of motion (EoM) for $x(r)$ using the Euler-Lagrange equation. Thus, it is found that
\begin{gather}
\mathcal{I}=\frac{w(r) f(r) \partial_r x}{\sqrt{1+f(r)\left(\partial_r x\right)^2}}, \quad w(r)=\frac{\mathrm{e}^{\mathrm{s} r^2}}{r^2}.
\end{gather}
$\mathcal{I}$ is a constant. We have $\partial _{r}x=\cot \alpha$ when $r=r_{v}$, and $\mathcal{I}$ can be expressed as
\begin{gather}
\mathcal{I}=\frac{w(r_{v}) f(r_{v}) \partial_r x}{\sqrt{1+f(r)\left(\partial_r x\right)^2}}, \quad w(r)=\frac{\mathrm{e}^{\mathrm{s} r^2}}{r^2}.
\end{gather}
Then $\partial_r x$  can be obtained:
\begin{gather}
\partial_r x=\sqrt{\frac{\omega\left(r_v\right)^2 f\left(r_v\right)^2}{\left(f\left(r_\nu\right)+\tan ^2 \alpha\right) \omega(r)^2 f(r)^2-f(r) w\left(r_v\right)^2 f\left(r_v\right)^2}}.
\end{gather}
Using equation $(10)$, we can obtain an expression for the separation distance $L$,
\begin{gather}
L=2 \int_0^{r_v} \frac{d x}{d r} d r.
\end{gather}
Next, we calculate the potential energy of doubly-heavy tetraquark. The energy of string $(1)$ can be got from the first item in the equation $(7)$,
\begin{gather}
E_1=\dfrac{S}{T}=\mathbf{g}\int_0^{r_s}\dfrac{\mathrm{e}^{sr^2}}{r^2}dr\sqrt{1+f(r)\left(\partial_r x\right)^2}.
\end{gather}
This expression is not well-defined, because
the integral diverges at $r = 0$. So we should subtract the divergent term
\begin{gather}
E_1=\dfrac{S}{T}=\mathbf{g}\int_{0}^{r_{v}}(\dfrac{1}{r^{2}}\mathrm{e}^{\mathbf{sr}^{2}}\sqrt{1+f(r)\left(\partial_{r}x\right)^{2}}-\dfrac{1}{r^{2}})dr-\dfrac{\mathbf{g}}{r_{v}}+c,
\end{gather}
here $c$ is a normalization constant. We can fix the constant by fitting the lattice results. For the heavy quarkonium, we can take 2c to fit the potential of heavy quarkonium with lattice. The choice of the normalization constant c for the energy of a single baryon configuration is equal to 3c.\cite{Andreev:2016some} String $(2)$ is calculated in the same way as string $(1)$, and thus $E_{1}=E_{2}$. String $(3)$ is described by the second term in equation $(7)$, and represents a straight string stretched between the baryon vertex and antibaryon vertex. The energy can be calculated as
\begin{gather}
E_3=\frac{S}{T}=\mathrm{g} \int_{r_v}^{r_{\bar{v}}} \frac{\mathrm{e}^{s r^2}}{r^2} d r.
\end{gather}
String $(4)$ and string $(5)$ are both described by the third term in equation $(7)$, and represent straight strings.
\begin{gather}
E_{4}=E_{5} =\dfrac{S}{T}=\int_{r_{\bar{v} } }^{r_{\bar{q}}  }\frac{e^{sr^{2} } }{r^{2} } \mathrm{d}r.
\end{gather}
Then, we  can get the energy of $QQ\bar{q}\bar{q}$ for this configuration.
\begin{gather}
\begin{aligned}
E_{Q Q \bar{q} \bar{q}}= & g\left(2 \int_0^{r_v}\left(\frac{e^{s r^2}}{r^2} \sqrt{1+f(r)\left(\partial_r x\right)^2}-\frac{1}{r^2}\right) \mathrm{d} r-\frac{2}{r_v}+\int_{r_v}^{r_{\bar{v}}} \frac{e^{s r^2}}{r^2} \mathrm{~d} r+2 \int_{r_{\bar{v}}}^{r_{\bar{q}}} \frac{e^{s r^2}}{r^2} \mathrm{~d} r\right. \\
& \left.+3 k \frac{e^{-2 s r_v^2}}{r_v} \sqrt{f\left(r_v\right)}+3 k \frac{e^{-2 s r_{\bar{v}}^2}}{r_{\bar{v}}} \sqrt{f\left(r_{\bar{v}}\right)}+2 n \frac{e^{\frac{1}{2} s r_{\bar{q}}^2}}{r_{\bar{q}}} \sqrt{f(r \bar{q})}\right)+2 c .
\end{aligned}
\end{gather}

It can be observed from equation (16) that energy is a function of $r_v$, $\alpha$, and $r_h$. To obtain the energy of this configuration, we will solve the position of the light antiquark. One crucial condition for equilibrium is that the net forces exerted on the vertices and antiquarks must cancel out. According to the model, the force (shown in Fig.\ref{fig1}) balance equation in the $r$ direction at $r_{\bar{q}}$ is given by:
\begin{gather}
2f_{\bar{q}}+e_{4}^{'}+e_{5}^{'}=0.
\end{gather}
Here $e_i$ is  the string tension \cite{Andreev:2021bfg}. By varying the action with respect to $r_{\bar{q}}$, we get the force  $f_{\bar{q}}=\left(0,-\mathbf{g}n\partial_{r_{\bar{q}}}(\frac{\mathrm{e}^{\frac{1}{2}\text{s}\mathrm{r}_{\bar{q}}^2}}{r_{\bar{q}}}\sqrt{f(r_{\bar{q}})})\right)$, $e_{4}^{'}=e_{5}^{'}=\textbf{g}w\left(r_{\bar{q}}\right)\left(0,-1\right)$ on the antiquark. Hence, the  equation $(17)$ becomes
\begin{gather}
-2nsr_{\bar{q}}^{2}(f^\frac{3}{2}(r_{\bar{q}}))+2n(f^\frac{3}{2}(r_{\bar{q}}))-nr_{\bar{q}}f' (r_{\bar{q}})-2f(r_{\bar{q}})e^\frac{sr_{\bar{q}}^2}{2}=0.
\end{gather}
 $r_{\bar{q}}$ (the position of the light antiquark) is only a function of $r_h$. This equation gives us the position  $r_{\bar{q}}$ at a fixed temperature.
Then, at $r_{\bar{v}}$, the force balance equation is
\begin{gather}
f_{\bar{v}}+e_3^{\prime}+e_4+e_5=0.
\end{gather}
Here $f_{\bar{v}}$ is the force on the  antibaryon vertex,  and each force is determined by
\begin{gather}
\begin{aligned}
f_{\bar{v}} & =\left(0,-3 \mathrm{~g} k \partial_{r_{\bar{v}}}\left(\frac{\mathrm{e}^{-2 s r_{\bar{v}}^2}}{r_{\bar{v}}} \sqrt{f\left(r_{\bar{v}}\right)}\right)\right) ,\\
e_3^{\prime} & =g w\left(r_{\bar{v}}\right)(0,-1), \\
e_4 & =e_5=g w\left(r_{\bar{v}}\right)(0,1) .\nonumber
\end{aligned}
\end{gather}
Then, the force balance equation changes
\begin{gather}
\left(4 s r_{\bar{v}}^2+1\right) k \sqrt{f\left(r_{\bar{v}}\right)} e^{-3 s r_{\bar{v}}^2}-\frac{1}{2} k r e^{-3 s r_{\bar{v}}^2} f^{\prime}\left(r_{\bar{v}}\right)+\frac{1}{3} \sqrt{f\left(r_{\bar{v}}\right)}=0 .
\end{gather}
At fixed temperature,  $r_{\bar{v}}$ can be determined via the above equation. At $r_{v}$, the force balance equation is
\begin{gather}
f_{v}  +e_{1}+e_{2}+e_{3} =0,
\end{gather}
 here the force on the baryon vertex is $f_v=\left(0,-3\mathbf{g}k\partial_{r_{v}}(\frac{\mathrm{e}^{-2s r_{v}^{2}}}{r_{v}}\sqrt{f(r_{v})})\right) $, and the string tensions at the ${r_{v}}$ are $e_{3}=\text{g}w\left(r_v\right)\left(0,1\right) $, $e_{1}=\mathbf{g}w\big(r_{v}\big)\big(-\frac{f(r_{v})}{\sqrt{t a n^{2}\alpha+f(r_{v})}},-\frac{1}{\sqrt{f(r_{v})c o t^{2}\alpha+1}}\big)$,
 $\begin{array}{r c l}{e_{2}}&{=}&{\mathbf{g}w\left(r_{v}\right)\left(\frac{f(r_{v})}{\sqrt{t a n^{2}\alpha+f(r_{v})}},-\frac{1}{\sqrt{f(r_{v})c o t^{2}\alpha+1}}\right)}\end{array}$. $\alpha$ is the  angle shown in Fig.\ref{fig1}. Putting these forces into equations $(21)$,  the force balance equation becomes
\begin{gather}
(4sr_{v}^2+1)k\sqrt{f(r_{v})} e^{-3sr_{v}^2}-\frac{1}{2}kre^{-3sr_{v}^2}f' (r_{v})+\frac{1}{3}\sqrt{f(r_{v})}(1-\frac{2}{\sqrt{f\left(r_v\right) \cot ^2 \alpha+1}})=0.
\end{gather}
By taking the variation of the action with respect to $r_{v}$, we can deduce the relationship between $r_{v}$ and $\alpha$.

\subsection{Slightly larger L}
\begin{figure}
    \centering
    \includegraphics[width=10cm]{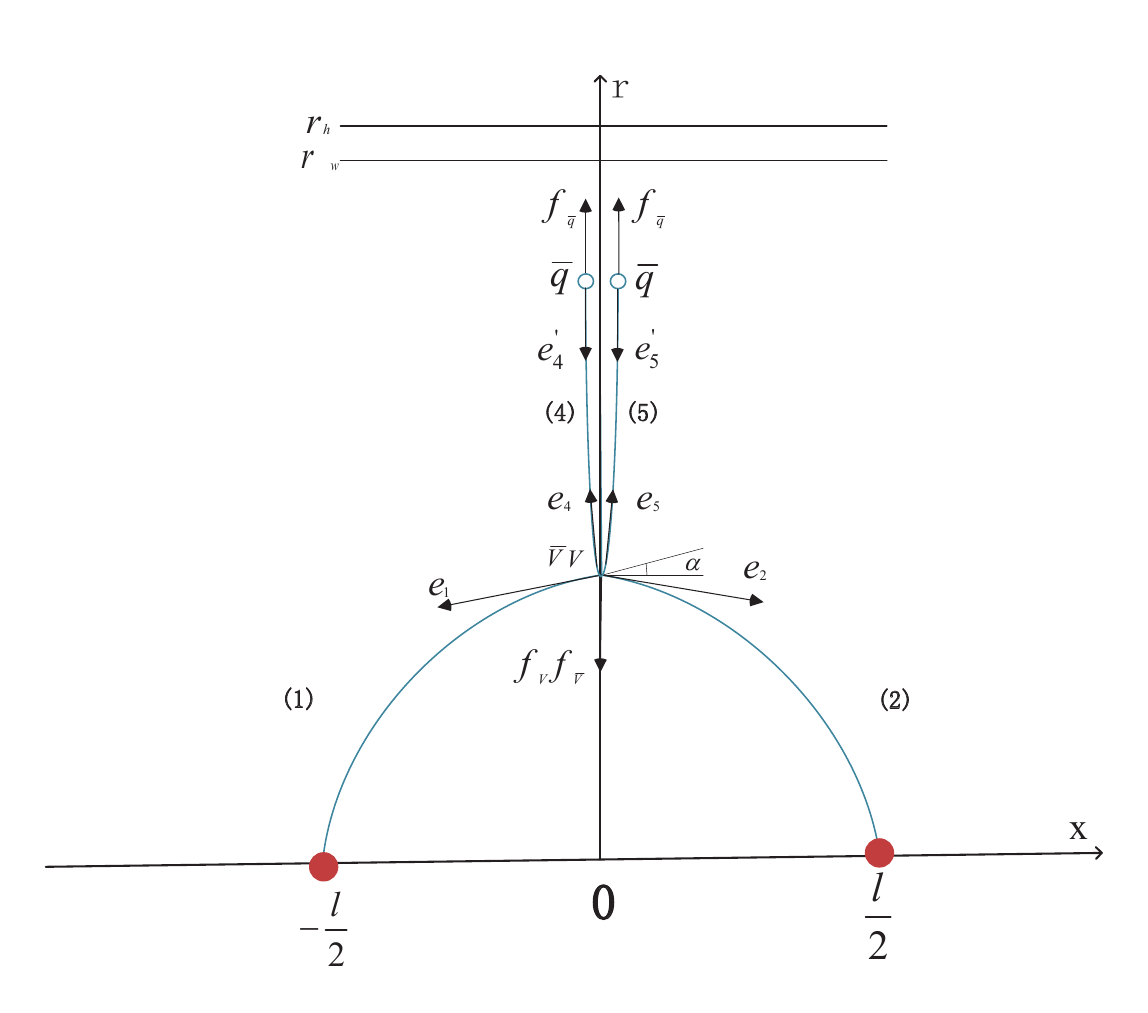}
    \caption{\label{fig2}A static configuration is formed by a slightly larger separation distance of a heavy-quark pair. The baryon vertex and the antibaryon vertex are in the same position. The force acting on the point is indicated by the black arrow.}
\end{figure}

The second configuration is shown in Fig.\ref{fig2}, and we consider the second configuration as one in which a string (3) contracts to a single point. The total action is now expressed by

\begin{gather}
S=\sum_{i=1, i \neq 3}^5 S_{\mathrm{NG}}^{(i)}+2 S_{\mathrm{v}}+2 S_{\mathrm{\bar{q}}}.
\end{gather}
We choose the same static gauge as before and the boundary conditions are
\begin{gather}
x^{(1)}(0)=-\frac{1}{2} \ell, \quad x^{(2)}(0)=\frac{1}{2} \ell, \quad x^{(i)}\left(r_v\right)=x^{(4,5)}\left(r_{\bar{q}}\right)=0.
\end{gather}

The separation distance of the slightly larger $L$ is still determined by equation $(11)$, which is the same as that used for the small $L$. Compared to the first configuration, the string (3) in the second configuration contracts to a point. Therefore, we need to consider other string tensions that satisfy equations $(13)-(15)$. Then, the energy of the slightly larger $L$ can be expressed as
\begin{gather}
\begin{aligned}
E_{Q Q \bar{q} \bar{q}}= & g\left(2 \int_0^{r_v}\left(\frac{e^{s r^2}}{r^2} \sqrt{1+f(r)\left(\partial_r x\right)^2}-\frac{1}{r^2}\right) \mathrm{d} r-\frac{2}{r_v}+2 \int_{r_v}^{r_{\bar{q}}} \frac{e^{s r^2}}{r^2} \mathrm{d} r\right. \\
& \left.+3 k \frac{e^{-2 s r_v^2}}{r_v} \sqrt{f\left(r_v\right)}+3 k \frac{e^{-2 s r_{\bar{v}}^2}}{r_{\bar{v}}} \sqrt{f\left(r_{\bar{v}}\right)}+2 n \frac{e^{\frac{1}{2} s r_{\bar{q}}^2}}{r_{\bar{q}}} \sqrt{f(r \bar{q})}\right)+2 c .
\end{aligned}
\end{gather}

As before, we will now proceed with solving the force balance equation. The location of $r_{\bar{q}}$ can be determined using equations (18). Then, the force balance equation at the point $r = r_v=r_{\bar{v}}$ is
\begin{gather}
f_{v}+f_{\bar{v}}+e_{1}+e_{2}+e_{4}+e_{5} =0.
\end{gather}
Each force is determined by
\begin{gather}
\begin{aligned}
f_v&=f_{\bar{v}}=\left(0,-3 g k\partial_{r_{v}}(\frac{\mathrm{e}^{-2s r_{v}^{2}}}{r_{v}}\sqrt{f(r_{v})})\right),\\
e_{1}&=g w_{(r_{v})}\big(-\frac{f(r_{v})}{\sqrt{t a n^{2}\alpha+f(r_{v})}},-\frac{1}{\sqrt{f(r_{v})c o t^{2}\alpha+1}}\big), \\
e_{2}&={g w_{(r_{v})}\left(\frac{f(r_{v})}{\sqrt{t a n^{2}\alpha+f(r_{v})}},-\frac{1}{\sqrt{f(r_{v})c o t^{2}\alpha+1}}\right)}, \\
e_{4}&=e_{5}=\text{g}w\left(r_v\right)\left(0,1\right).\nonumber
\end{aligned}
\end{gather}
Then, the equation $(26)$ becomes
\begin{gather}
(24sr_{v}^2f(r_{v})+6f(r_{v})-3rf'(r_{v}))ke^{-3sr_{v}^2}-\frac{2}{\sqrt{f\left(r_v\right) \cot ^2 \alpha+1}}+2\sqrt{f(r_v)}=0.
\end{gather}

\subsection{Intermediate L}

\begin{figure}
    \centering
    \includegraphics[width=10cm]{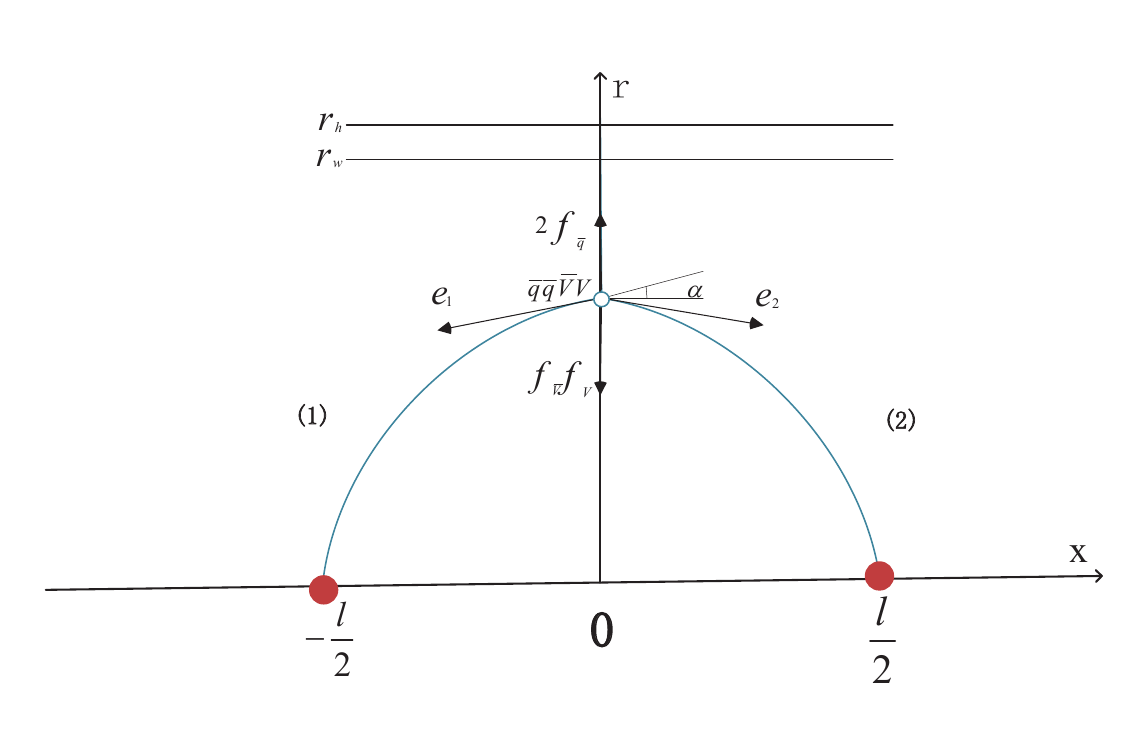}
    \caption{\label{fig3} Static configuration at an intermediate separation distance of a heavy-quark pair. The baryon vertex, antiquark and the antibaryon vertex are in the same position. The force acting on the point is indicated by the black arrow.}
\end{figure}

The third configuration, as shown in Fig.\ref{fig3}, is characterized by the compression of points $r_{\bar{q}}$, $r_{\bar{v}}$ and $r_v$ into a single location. The configuration of the total action is given by
\begin{gather}
S=\sum_{i=1}^2 S_{\mathrm{NG}}^{(i)}+2 S_{\mathrm{v}}+2 S_{\mathrm{\bar{q}}}.
\end{gather}
Choosing the static gauge in the Nambu-Goto actions as before, then  the boundary conditions can be obtained
\begin{gather}
x^{(1)}(0)=-\frac{1}{2} \ell, \quad x^{(2)}(0)=\frac{1}{2} \ell, \quad x^{(i)}\left(r_v\right)=0.
\end{gather}
Equation $(11)$ gives us the separation distance in this configuration. To calculate the energies, we only need to consider the range from $r=0$ to $r=r_v$, as specified by equations $(13)$. Naturally, the energy of the configuration is
\begin{gather}
\begin{aligned}
E_{Q Q \bar{q} \bar{q}}= & g\left(2 \int_0^{r_v}\left(\frac{e^{s r^2}}{r^2} \sqrt{1+f(r)\left(\partial_r x\right)^2}-\frac{1}{r^2}\right) \mathrm{d} r-\frac{2}{r_v}+6 k \frac{e^{-2 s r_{\bar{v}}{ }^2}}{r_{\bar{v}}} \sqrt{f\left(r_{\bar{v})}\right.}\right. \\
& \left.+2 n \frac{e^{\frac{1}{2} s r_{\bar{q}}^2}}{r_{\bar{q}}} \sqrt{f(r \bar{q})}\right)+2 c .
\end{aligned}
\end{gather}
The force balance equation at the point
$r=r_v= r_{\bar{v}}= r_{\bar{q}}$ is
\begin{gather}
\begin{aligned}
f_{v}+f_{\bar{v}}+2f_{\bar{q}}+e_{1}+e_{2} =0.
\end{aligned}
\end{gather}
Each force is determined by
\begin{gather}
\begin{aligned}
f_v&=f_{\bar{v}}=\left(0,-3 g k\partial_{r_{v}}(\frac{\mathrm{e}^{-2s r_{v}^{2}}}{r_{v}}\sqrt{f(r_{v})})\right), \\
e_{1}&=g w_{(r_{v})}\big(-\frac{f(r_{v})}{\sqrt{t a n^{2}\alpha+f(r_{v})}},-\frac{1}{\sqrt{f(r_{v})c o t^{2}\alpha+1}}\big),\\
e_{2}&={g w_{(r_{v})}\left(\frac{f(r_{v})}{\sqrt{t a n^{2}\alpha+f(r_{v})}},-\frac{1}{\sqrt{f(r_{v})c o t^{2}\alpha+1}}\right)}, \\
{f}_{\bar{q}}&=\left(0,-\mathbf{g}n\partial_{r_{\bar{q}}}(\frac{\mathrm{e}^{\frac{1}{2}\text{s}\mathrm{r}_{\bar{q}}^2}}{r_{\bar{q}}}\sqrt{f(r_{\bar{q}})})\right).\nonumber
\end{aligned}
\end{gather}
Here $r_{\bar{q}}=r_{\bar{v}}=r_{v}$, then, the equation $(31)$ becomes
\begin{gather}
\begin{aligned}
&ke^{-2sr_{v}}(24sr_{v}^2f(r_{v})+6f(r_{v})-3r_{v}f' (r_{v}))+ne^{\frac{1}{2}sr_{v}^2}(2f(r_{v})
-2sr_{v}^2f(r_{v})\\
&-r_{v}f'(r_{v}))-\frac{2\sqrt{f(r)} e^{sr_{v}^2}}{\sqrt{f\left(r_v\right) \cot ^2 \alpha+1}}=0.
\end{aligned}
\end{gather}

\subsection{Large L}

\begin{figure}
    \centering
    \includegraphics[width=10cm]{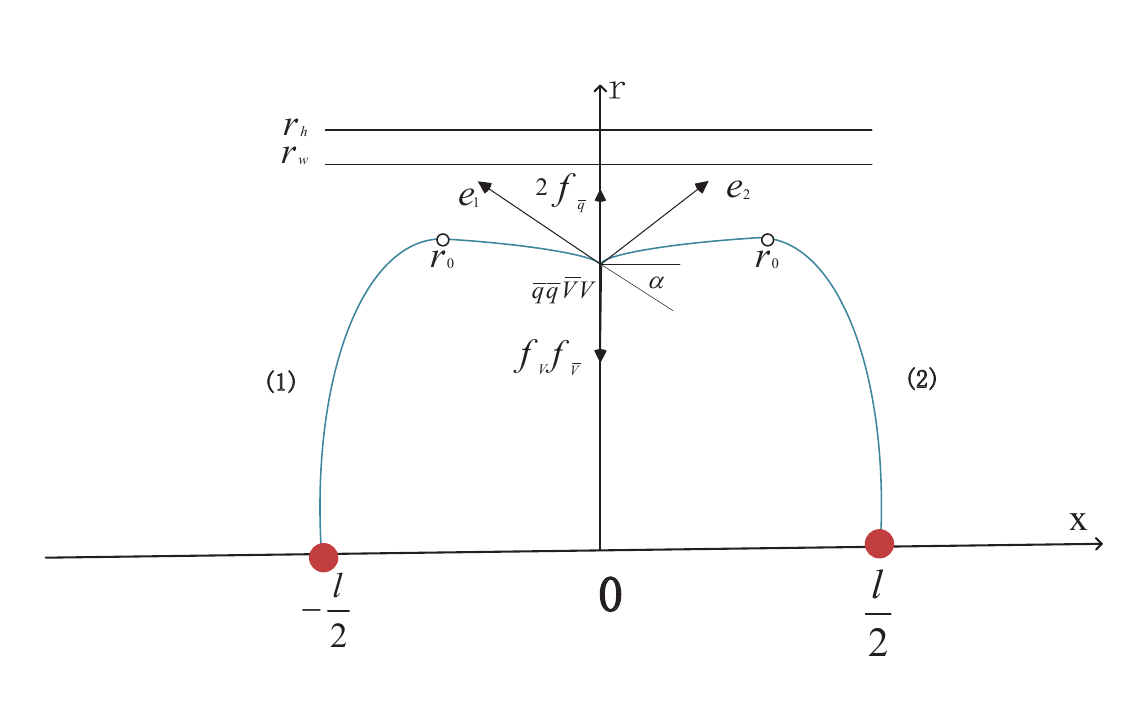}
    \caption{\label{fig4} Static configuration at a large separation distance of a heavy-quark pair. There are two turning points at strings $(1)$ and $(2)$ and they are symmetric about the Y-axis. The force acting on the point is indicated by the black arrow.}
\end{figure}
Fig.\ref{fig4}  illustrates the fourth configuration. The total action for this configuration, denoted as intermediate $L$, is given by equation $(28)$. We choose another  static gauge $\xi^1=t$ and $\xi^2=x$ in the Nambu-Goto actions and the boundary conditions are
\begin{gather}
\begin{aligned}
r^{(1)}(-L/2)=r^{(2)}(L/2)=0,\quad r^{(i)}(0)=r_v.
\end{aligned}
\end{gather}
Then, the total action becomes
\begin{gather}
\begin{aligned}
S=&gT\Big (\int_{-L/2}^{0}\frac{e^{sr^{2}}}{r^{2}}\sqrt{f(r)+(\partial_{r}x)^2\mathrm{d}x}+\int_{0}^{L/2}\frac{e^{sr^{2}}}{r^{2}}\sqrt{f(r)+(\partial_{r}x)^2\mathrm{d}x}\\&+6k\frac{e^{-2sr_{v}^{2}}}{r_{v}}\sqrt{f(r_{v})}+2n\frac{e^{-\frac{1}{2} sr_{\bar{q} }^{2}}}{r_{\bar{q}}}\sqrt{f(r_{\bar{q}})}\Big ).
\end{aligned}
\end{gather}
The action for string $(1)$ is given by the first term in equation $(34)$. The subsequent step is to compute the first integral.
\begin{gather}
\begin{aligned}
\mathcal{I=}\dfrac{w(r)f(r)}{\sqrt{f(r)+\left(\partial_x r\right)^2}}.
\end{aligned}
\end{gather}
$\mathcal{I}$ is a constant. At the $r_0$ and $r_v$ points, we can obtain, respectively
\begin{gather}
\begin{aligned}
\dfrac{w(r)f(r)}{\sqrt{f(r)+\left(\partial_x r\right)^2}}=w(r_0)\sqrt{f(r_0)}.
\end{aligned}
\end{gather}
\begin{gather}
\begin{aligned}
\frac{w(r_v)f(r_v)}{\sqrt{f(r_v)+tan\alpha^2}}=w(r_0)\sqrt{f(r_0)}.
\end{aligned}
\end{gather}
Here $\partial_x r$  can be obtained from equation $(36)$,$(37)$
\begin{gather}
\begin{aligned}
\partial_x r=\sqrt{\frac{w(r)^2f(r)^2f(r_0)-f(r)w(r_0)^2f(r_0)^2}{w(r_0)^2f(r_0)^2}}.
\end{aligned}
\end{gather}
The large $L$ configuration has a turning point at $r_{0}$, so the distance between heavy quarks is calculated in two parts. Then, the separation distance can be expressed as
\begin{gather}
\begin{aligned}
L=2(L_1+L_2)=2(\int_0^{r_0}\dfrac{1}{r'}dr+\int_{r_v}^{r_0}\dfrac{1}{r'}dr).
\end{aligned}
\end{gather}
Here $r'$ denotes $\partial_x r$. By substituting equation $(38)$ into equation $(39)$, the separation distance can be obtained. The energy of string $(1)$ was calculated in two parts, which is similar to the separation distance. Therefore, the energy of string $(1)$ is given by
\begin{gather}
\begin{aligned}
E_1=E_{R_1}+E_{R_2}=\mathbf{g}\int_0^{r_0}w(r)\sqrt{1+f(r)x'^2}dr+\mathbf{g}\int_{r_{\bar{q}}}^{r_0}w(r)\sqrt{1+f(r)x'^2}dr.
\end{aligned}
\end{gather}
Same as the case of small $L$, we also need to subtract the divergent term here. Thus, equation $(40)$ can be expressed as
\begin{gather}
\begin{aligned}
E_1=\mathbf{g}\int_0^{r_0}(w(r)\sqrt{1+f(r)x'^2}-\dfrac{1}{r^2})dr+\mathbf{g}\int_{r_v}^{r_0}w(r)\sqrt{1+f(r)x'^2}dr-\dfrac{1}{r_0}+2c.
\end{aligned}
\end{gather}
String (2) is calculated in the same way as string (1). Therefore, the total energy of the configuration is
\begin{gather}
\begin{aligned}
E_{Q Q \bar{q} \bar{q}}= & \mathrm{2g} \int_0^{r_0}(w(r) \sqrt{1+f(r) x^{\prime 2}}-\frac{1}{r^2}) d r+2 \mathrm{~g} \int_{r_v}^{r_0} w(r) \sqrt{1+f(r) x^{\prime 2}} d r-\frac{2g}{r_0}
\\&+6 k g \frac{e^{-2 s r_{\bar{v}}}}{r_{\bar{v}}} \sqrt{f(r_{\bar{v}})}+2 n g \frac{e^{\frac{1}{2} s r_{\bar{q}}^2}}{r_{\bar{q}}} \sqrt{f(r_{\bar{q}})}+2 c .
\end{aligned}
\end{gather}

The force balance equation is the same as that for intermediate $L$, and the expressions of force are also the same as intermediate $L$.  Equation $(37)$ represents the functional relationship between $r_{v}$ and $\alpha$ when the temperature is fixed. We can first solve equations (32) and (37), and then substitute the numerical values into equations (39) and (42) to obtain the solutions for $L$ and $E_{QQ\bar{q}\bar{q}}$. For $\alpha$, at small separate distance of heavy quarks, as $r_v$ increases, the distance between the heavy quarks increases, and the curvature of the string between heavy quarks increases, causing $\alpha$  to decrease. When the string reaches $\alpha=0$ and $r_v$ increases further, the string will become an M-shape, at which point $\alpha$ becomes negative.

\section{Numerical results and discussion}\label{sec:04}
\subsection{T=0.08 GeV}

\begin{figure}
	\centering
	\includegraphics[width=8.5cm]{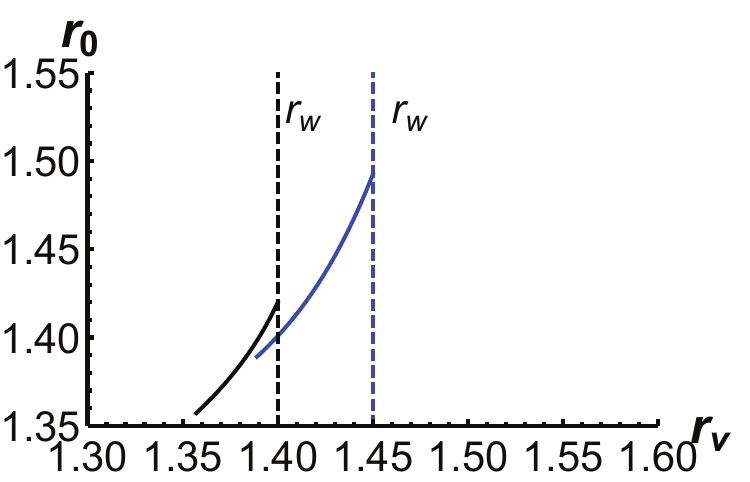}
	\caption{\label{fig5} $r_0$ as a function of $r_v$ in the larger $L$ configuration, where the black line is $r_0$ at $T = 0$, while the blue line is $r_0$ at $T = 0.08$ $\rm{GeV}$. $r_w$ indicates the position of a dynamic wall.}
\end{figure}

At a low temperature of $0.08$ $\rm{GeV}$, the configurations of $\rm{QQ\bar{q}\bar{q}}$ are confined. Below the black-hole horizon, there exists a dynamic wall  that prevents  the string from crossing.  In this part, we investigate four configurations of $\rm{QQ\bar{q}\bar{q}}$ at this temperature.

First, we will give a discussion about the dynamic wall at finite temperature. Different from the quark-antiquark pair, the configuration of QQqq shows a "M" shape at large separation distance. Thus, the maximum value of $r_0$ at infinite separation distance gives the position of dynamic wall.  By calculating the position of the dynamic wall, we obtain $r_{w}=1.40$$\rm{GeV^{-1}}$ at $T = 0$ and $r_{w}=1.45$$\rm{GeV^{-1}}$ at $T = 0.08$ $\rm{GeV}$ as shown in Fig.\ref{fig5}. The increase of temperature leads to an increase of the position of dynamic wall in the $r$ direction.

\begin{figure}
    \centering
    \includegraphics[width=8.5cm]{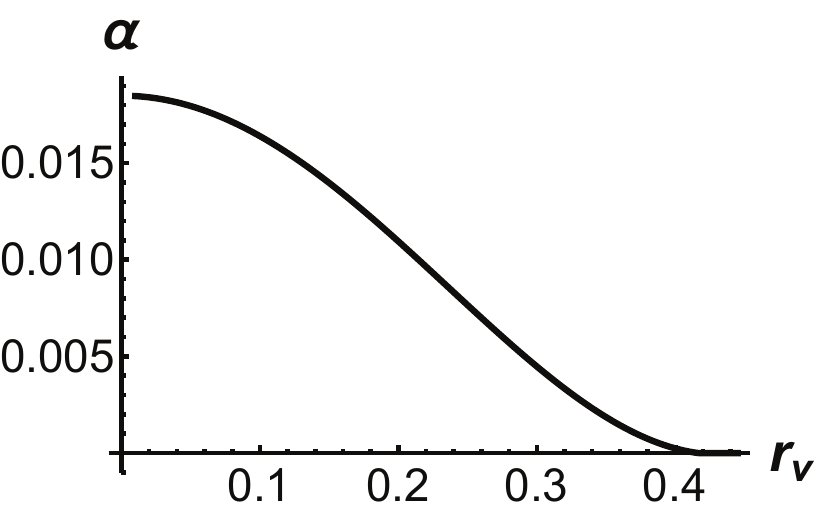}
    \caption{\label{fig6} $\alpha$ as a function of $r_v$ in the small $L$ configuration. The unit for $r_v$ is  $\rm{GeV^{-1}}$.}
\end{figure}

For small $L$, we use equation $(18)$ to calculate the position of the antiquark.  The result shows that at a temperature of $0.08$ $\rm{GeV}$, $r_{\bar{q}}=1.13$ $\rm{GeV^{-1}}$. Subsequently, the antibaryon vertex position can be calculated using equation $(20)$, which yields $r_{\bar{v}}=0.445$ $\rm{GeV^{-1}}$. Within the range of $0<r<r_{v}$, we can calculate $\alpha$ using equation $(22)$, as illustrated in Fig.\ref{fig6}. It can be observed that $\alpha$ exhibits a decreasing trend as $r_v$ increases. Then, the separation distance of a heavy-quark pair and its corresponding energy can then be obtained using equations $(11)$ and $(16)$, respectively. These results are presented in Fig.\ref{fig10} and Fig.\ref{fig11}.

\begin{figure}
    \centering
    \includegraphics[width=8.5cm]{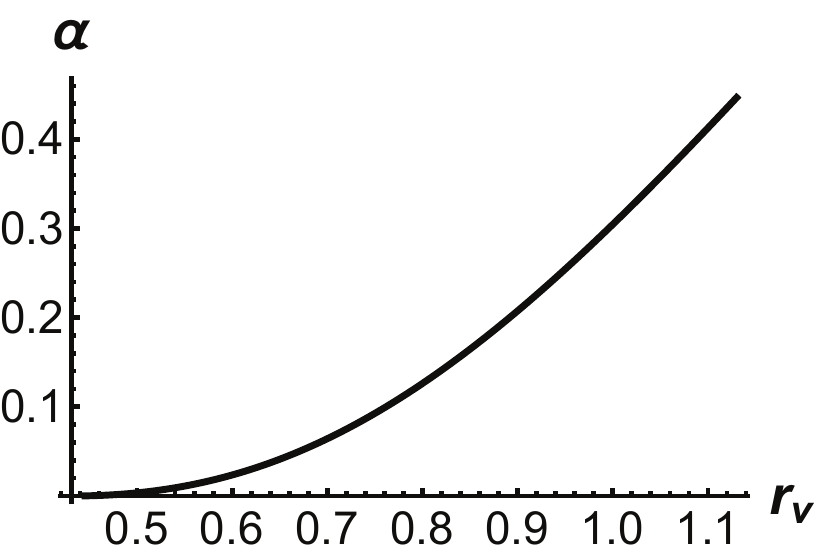}
    \caption{\label{fig7} $\alpha$ as a function of $r_v$ in the slightly larger $L$ configuration. The unit for $r_v$ is $\rm{GeV^{-1}}$.}
\end{figure}

When $r_v$ increases to $r_{\bar{v}}$, it reaches a critical value and transitions to the second configuration. In the second configuration, we can still determine the position of antiquark using equation $(18)$. Subsequently, we use equation $(27)$ to obtain $\alpha$, as depicted in Fig.\ref{fig7}. $\alpha$ exhibits an increasing trend as $r_v$ increases, reaching its peak when $r_v=r_{\bar{v}}$. Next, we can obtain the separation distance and energy of the second configuration using equations $(11)$ and $(25)$, and it is observed that as $r_v$ increases, $L$ and $E$ also exhibit an upward trend. When $r_v$ reaches $r_{\bar{v}}$, it attains its maximum value. Beyond this point, the configuration switches to the third configuration.

\begin{figure}
    \centering
    \includegraphics[width=8.5cm]{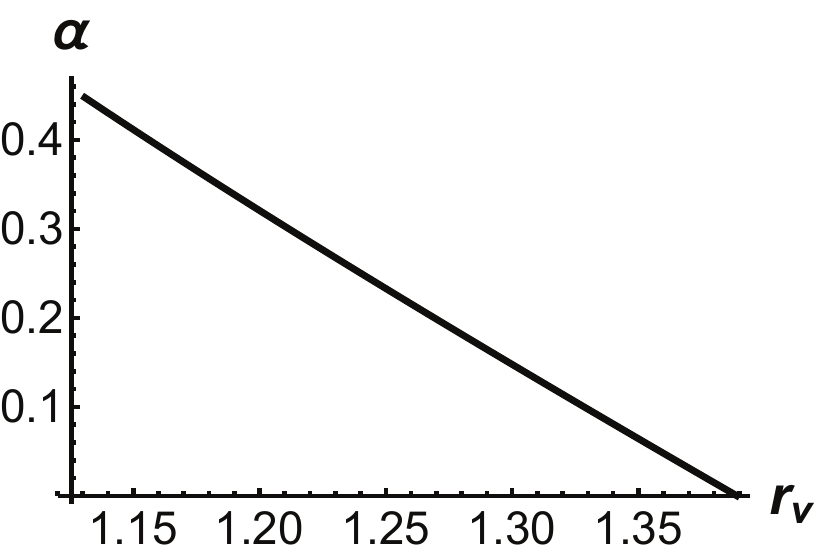}
    \caption{\label{fig8} $\alpha$ as a function of $r_v$ in the intermediate $L$ configuration. The unit for $r_v$ is $\rm{GeV^{-1}}$.}
\end{figure}

In the third configuration, $r_{\bar{v}}$ overlaps with the $r_{\bar{q}}$ point. Using equations $(32)$, we can establish a functional relationship between $r_v$ and $\alpha$, as shown in Fig.\ref{fig8}. Clearly, $\alpha$ exhibits a linear decrease with increasing $r_v$ until it reaches 0. $E$ and $L$ can be obtained using equations $(11)$ and $(30)$. In this configuration, $E$ varies linearly with $L$. When $r_v$ exceeds $r_{\bar{q}}$, the configuration shifts into the fourth configuration.

\begin{figure}
    \centering
    \includegraphics[width=8.5cm]{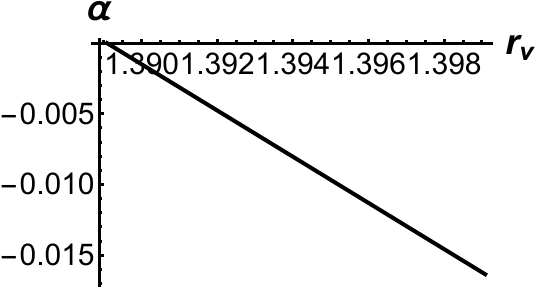}
    \caption{\label{fig9} $\alpha$ as a function of $r_v$ in the large $L$ configuration. The unit for $r_v$ is $\rm{GeV^{-1}}$. }
\end{figure}

\begin{figure}
    \centering
    \includegraphics[width=8.5cm]{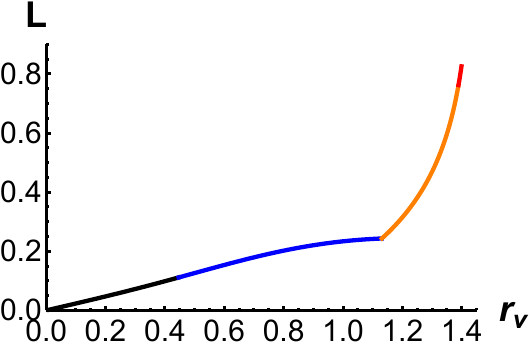}
    \caption{\label{fig10} Separation distance $L$ as a function of $r_v$, where the black line represents the configuration with small $L$, blue represents slightly larger $L$, orange represents intermediate $L$, and red represents larger $L$. The unit of $L$ is in $\rm{fm}$ and that of $r_v$ is in $\rm{GeV^{-1}}$.}
\end{figure}

\begin{figure}
    \centering
    \includegraphics[width=8.5cm]{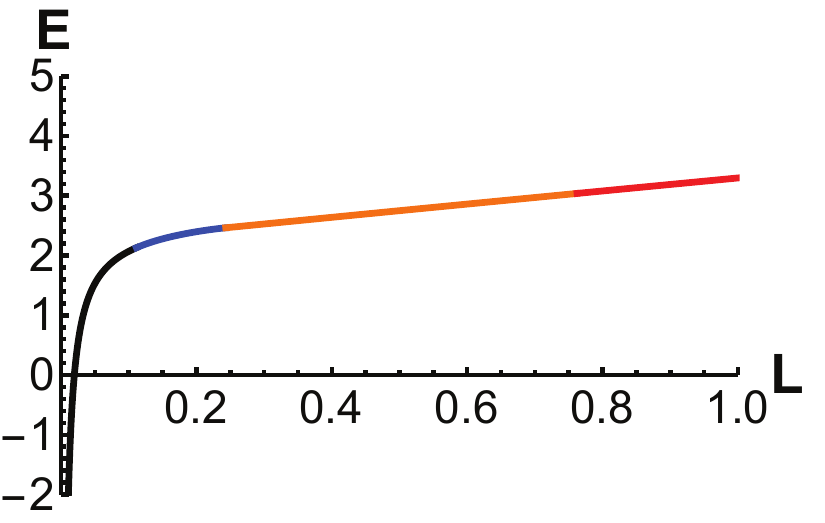}
    \caption{\label{fig11} Energy $E$ as a function of separation distance $L$  at $T=0.08$ $\rm{GeV}$. The black line represents the configuration with small $L$, blue line  represents slightly larger $L$, orange line represents intermediate $L$, and red line  represents larger $L$. The unit of $E$ is in $\rm{GeV}$ and that of $r_v$ is in $\rm{GeV^{-1}}$. }
\end{figure}

Then, in the configuration with large $L$, as $r_v$ increases below the dynamic wall, strings $(1)$ and $(2)$ exhibit turning points. $\alpha$ continues to decrease as $r_v$ increases, as shown in Fig.\ref{fig9}. However, unlike before, $\alpha$ becomes negative. As shown in Fig.\ref{fig9}, the maximum value of $r_v$ is $1.40$ $\rm{GeV^{-1}}$, which corresponds to the position of the dynamic wall  at $r_w\approx r_0\approx1.45$ $\rm{GeV^{-1}}$. We can obtain $E$ and $L$ using equations $(39)$ and $(42)$.

At $T=0.08$ $\rm{GeV}$, the separation distance and energy plots for the tetraquark configuration are illustrated in Fig.\ref{fig10} and Fig.\ref{fig11}, respectively. As can be observed from the figure, there is a smooth connection of $L$ for each configuration. As $r_v$ increases, $L$ also approaches infinity. This implies that $r_v$ cannot be infinite, and when the maximum value is exceeded, the configuration will decay into another state.  We will discuss this further in the upcoming section. The energy increases as the separation distance increases. Notably, the energy is dominated by Coulomb potential at small separation distances and linear potential at large distances.

\subsection{T=0.115GeV}

At a temperature of $T = 0.115$ $\rm{GeV}$, the configurations of $\rm{QQ\bar{q}\bar{q}}$ are deconfined. For a deconfined  $\rm{QQ\bar{q}\bar{q}}$, the melting of  $\rm{QQ\bar{q}\bar{q}}$ can happen at a certain separation distance. If the separation distance of heavy quark-anqitquark pair is small enough, the  $\rm{QQ\bar{q}\bar{q}}$ will not melt even at high temperature. When increasing the distance of heavy quark-anqitquark pair, the color screening becomes important and the  $\rm{QQ\bar{q}\bar{q}}$ will melt. In summary, the dynamic wall disappears and $\rm{QQ\bar{q}\bar{q}}$ will melt at a sufficiently far distance in this stage.

\begin{figure}
    \centering
    \includegraphics[width=8.5cm]{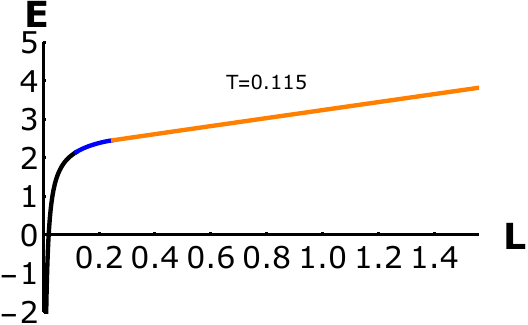}
    \caption{\label{fig12} Energy $E$ as a function of separation distance $L$  at $T=0.115$  $\rm{GeV}$. The black line represents the configuration with small $L$, blue line represents slightly larger $L$ and orange line represents intermediate $L$. The unit of $E$ is in $\rm{GeV}$ and that of $r_v$ is in $\rm{GeV^{-1}}$. }
\end{figure}

Firstly, similar to $T=0.08$ $\rm{GeV}$, we focus on the first configuration. The position of the antiquark can be determined using equation $(18)$, which yields $r_{\bar{q}}=1.1677$ $\rm{GeV^{-1}}$ when $T=0.115$ $\rm{GeV}$. Next, we can calculate the position of the antibaryon vertex using equation $(20)$, which yields $r_{\bar{v}}=0.4626$ $\rm{GeV^{-1}}$, and then use equation $(22)$ to determine $\alpha$.  $L$ and $E$ can still be obtained from equations $(11)$ and $(16)$, respectively. We can then proceed to investigate the second and third configurations using a similar approach as the $T=0.08$ $\rm{GeV}$ calculation. This will enable us to determine the separation distance and energy. As the value of $r_v$ continues to increase, the value of $L$ also exhibits a tendency to increase. However, as $L$ approaches a maximum value, it will tend towards infinity, indicating that the quark is now free and large $L$ configurations will no longer be possible as shown in Fig.\ref{fig12} \cite{Peeters:2006iu,Hoyos-Badajoz:2006dzi,Fadafan:2012qy}.
First, similar to $T=0.08$ $\rm{GeV}$, we focus on the first configuration. The position of the antiquark can be determined using equation $(18)$, which yields $r_{\bar{q}}=1.1677$ $\rm{GeV^{-1}}$ when $T=0.115$ $\rm{GeV}$. Next, we can calculate the position of the antibaryon vertex using equation $(20)$, which yields $r_{\bar{v}}=0.4626$ $\rm{GeV^{-1}}$, and then use equation $(22)$ to determine $\alpha$.  $L$ and $E$ can still be obtained from equations $(11)$ and $(16)$, respectively. Subsequently, we can investigate the second and third configurations at this temperature by adopting the same method as the $T=0.08$ $\rm{GeV}$ calculation, which allows us to determine the separation distance of heavy quarks and energy. As the value of $r_v$ continues to increase, the value of $L$ also exhibits a tendency to increase. However, as $L$ approaches a maximum value ($L_{max}=1.56\rm{fm}$), it will tend towards infinity, indicating that the quark is now free and large $L$ configurations will no longer be possible \cite{Peeters:2006iu,Hoyos-Badajoz:2006dzi,Fadafan:2012qy}. The diagram of energy and separation distance is shown in Fig.\ref{fig12}.

\begin{figure}
    \centering
    \includegraphics[width=8.5cm]{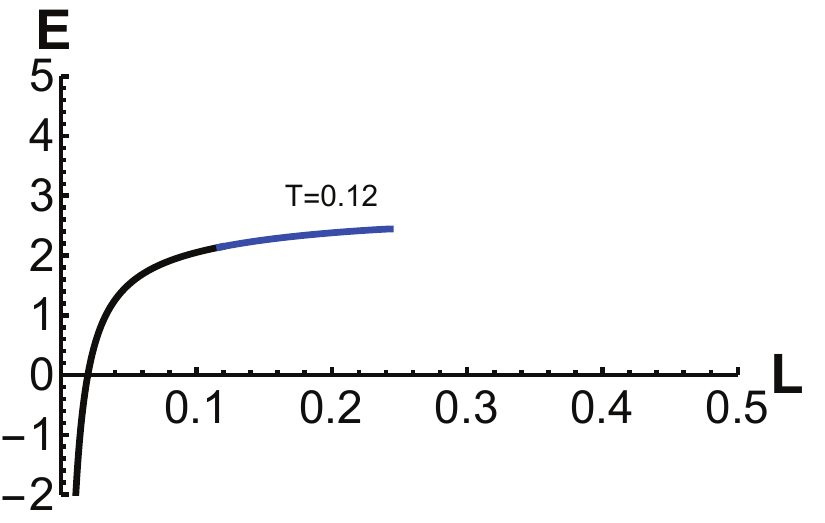}
    \caption{\label{fig13} Energy $E$ as a function of separation distance $L$ at $T=0.12$ $\rm{GeV}$. The black line represents the configuration with small $L$ and blue line represents slightly larger $L$. The unit of $E$ is in $\rm{GeV}$ and that of $r_v$ is in $\rm{GeV^{-1}}$.}
\end{figure}

\subsection{T=0.12GeV}

To begin with, we can use equation $(18)$ to determine the position of the antiquark, which is $r_{\bar{q}}=1.1785$ $\rm{GeV^{-1}}$ when $T = 0.12$ $\rm{GeV}$. Next, we can use equation $(20)$ to calculate the position of the antibaryon vertex, which yields $r_{\bar{v}}=0.46574$ $\rm{GeV^{-1}}$. Following this, we can similarly obtain functional expressions for $\alpha$, $L$, and $E$ as $r_v$ increased, which are illustrated in Fig.\ref{fig13}. As depicted in the figure, both $L$ and $E$ exhibit linear growth with respect to $r_v$, and the energy function $E$ shows a Cornell-like potential.  At slightly larger $L$,  equations $(11)$ and $(25)$ are employed to calculate the separation distance and energy. However, at this temperature, as $r_v$ keeps increasing, the configuration will eventually collapse, leading to the quark becoming a free state. Therefore, there exists a maximum value of $L$, beyond which $L$ will tend towards infinity. Consequently, large values of $L$ are not possible at this temperature.

\begin{figure}
    \centering
    \includegraphics[width=8.5cm]{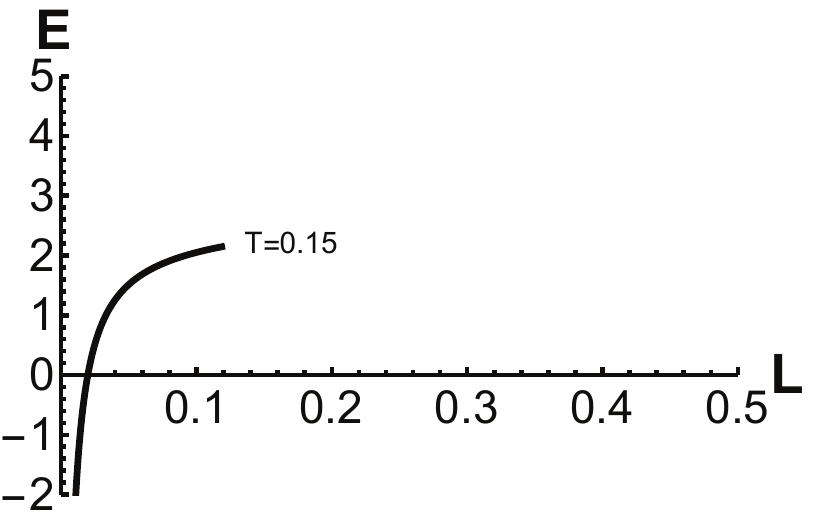}
    \caption{\label{fig14} Energy $E$ as a function of separation distance $L$ at $T=0.15$ $\rm{GeV}$. The black line represents the configuration with small $L$. The unit of $E$ is in $\rm{GeV}$ and that of $r_v$ is in $\rm{GeV^{-1}}$.}
\end{figure}

\subsection{T=0.15GeV}

At this temperature, using the force balance equation, the position of the quark or antibaryon vertex is determined to be $r_{\bar{v}}=0.4726$ $\rm{GeV^{-1}}$. However, we have discovered that at this temperature there is no solution for the position of the antiquark, $r_{\bar{q}}$. Therefore, we have computed the separation distance and energy for the first configuration, which are displayed in Fig.\ref{fig14}. As $r_{\bar{v}}$ surpasses a certain value, the $\rm{QQ\bar{q}\bar{q}}$ configuration collapses, resulting in the quark becoming free.

\begin{figure}
    \centering
    \includegraphics[width=8.5cm]{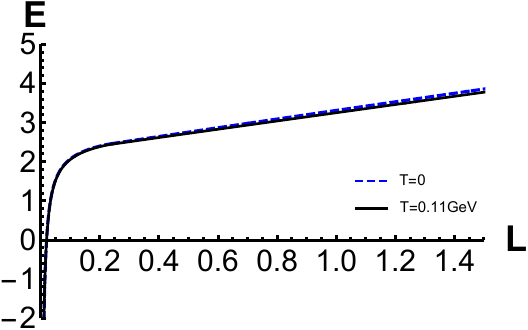}
    \caption{\label{fig15} The blue dash line is the energy at zero temperature, while the black line is the energy at $T=0.11$ $\rm{GeV}$.}
\end{figure}

\subsection{Short summary}

Based on the calculation results of the $\rm{QQ\bar{q}\bar{q}}$ potential at the above mentioned four temperatures, it shows that both $r_{\bar{q}}$ and $r_{\bar{v}}$ become larger and gradually approach the position of the dynamic wall as the temperature increases. Furthermore, the melting of the $\rm{QQ\bar{q}\bar{q}}$ configuration happens at a small distance as the temperature increases. Comparing the two lines for $T=0$ and $T=0.11$ $\rm{GeV}$ in Fig.\ref{fig15}, we observe that as temperature increases, the same separation distance $L$ corresponds to a lower energy value. Besides, at small distances, $\rm{QQ\bar{q}\bar{q}}$ exhibits Coulombic behavior, whereas at large distances, the behavior is linear  at finite temperature \cite{White:2007tu,Gribov:1977wm,Karch:2006pv}.

\section{decay modes }\label{sec:05}

\subsection{\texorpdfstring {$\rm{QQ\bar{q}\bar{q}}$}{} }

\begin{figure}
    \centering
    \includegraphics[width=15cm]{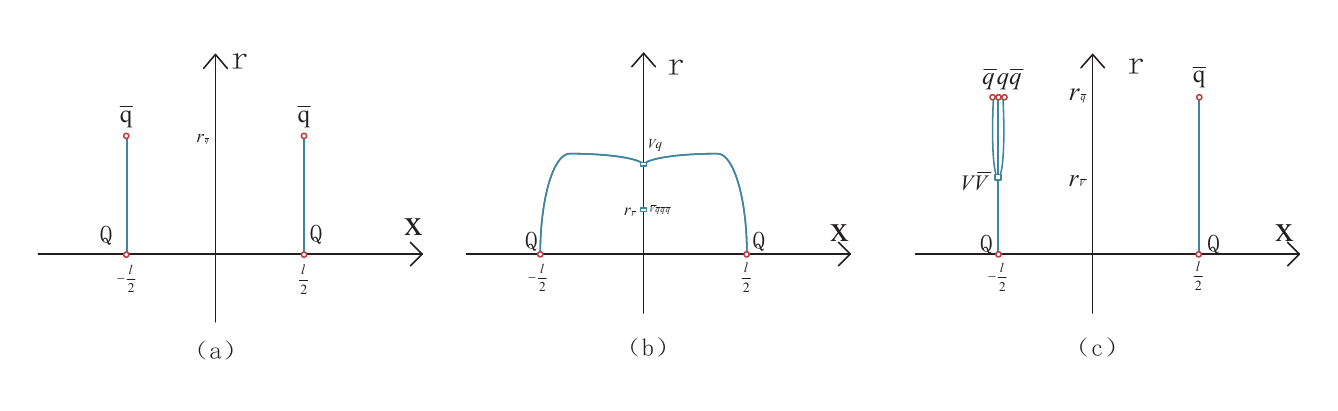}    
    \caption{\label{fig16}Three disconneccted configuration of $\rm{QQ\bar{q}\bar{q}}$.}
\end{figure}

During the confinement phase, quarks are confined within hadrons. However, as the distance between heavy quarks increases, the strings connecting them eventually break \cite{Bazavov:2011nk}. In this context, we shall consider three decay modes for $\rm{QQ\bar{q}\bar{q}}$ configurations.
\begin{gather}
\begin{aligned}
\begin{matrix}
&\nearrow Q\bar{q}+Q\bar{q}\\
\\QQ\bar{q}\bar{q}& \to QQq+\bar{q}\bar{q}\bar{q}\\ \\ &\searrow Qq\bar{q}\bar{q}+Q\bar{q}.
\end{matrix}
\end{aligned}
\end{gather}

Fig.\ref{fig16} displays the configuration diagram for the three possible decay modes. We will proceed to analyze the energy of each disconnected configuration. $\rm{Q\bar{q}}$ consists of a fundamental string and an antiquark, with the total action given by $S_{Q\bar{q}}=S_{\mathrm{NG}}+S_{\mathrm{q}}$. $\rm{QQq}$ comprises two strings, a vertex, and a light quark. The total action can be written as $S_{QQq}=\sum_{i=1}^2S_{\text{NG}}^{(i)}+S_{\text{vert}}+S_{{\bar{q}}}$. On the other hand, $\rm{Qq\bar{q}\bar{q}}$ consists of four strings, a vertex, a antibaryon vertex, a light quark and two antiquarks. The total action is $S_{Qq\bar{q}\bar{q}}=\sum_{i=1}^4S_{\text{NG}}^{(i)}+S_{v}+S_{\bar{v}}+2S_{{\bar{q}}}+S_{q}$. Furthermore, $\bar{q}\bar{q}\bar{q}$ consists of a antibaryon vertex and three antiquarks, with the total action $S_{\bar{q}\bar{q}\bar{q}}=S_{\bar{v}}+3S_{{\bar{q}}}$. We employ the same static gauge as before and obtain the total action for each configuration by specifying appropriate boundary conditions. The detailed calculation process of the energy of each configuration is given in Refs. \cite{Chen3kuake,Andreev:2021bfg,Andreev:2016some,Andreev3kuake}.  Subsequently, The action of each decay is then calculated.
\begin{gather}
\begin{aligned}
S_{1}&=S_{Q\bar{q}}+S_{Q\bar{q}}
\\S_{2}&=S_{QQq}+S_{\bar{q}\bar{q}\bar{q}}
\\S_{3}&=S_{Qq\bar{q}\bar{q}}+S_{Q\bar{q}}.
\end{aligned}
\end{gather}
\begin{figure}
    \centering
    \includegraphics[width=15cm]{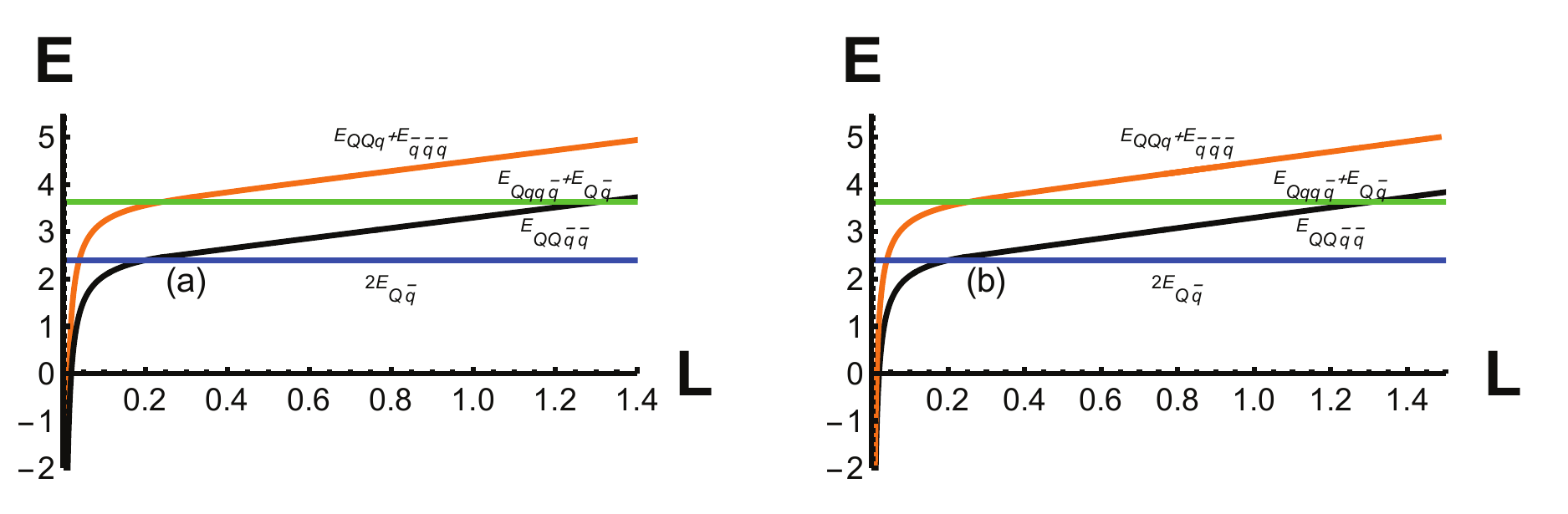}
\caption{\label{fig17} (a)$\rm{QQ\bar{q}\bar{q}}$ and his three disconnected configuration energy at temperature $T=0$ $\rm{GeV}$. (b)$\rm{QQ\bar{q}\bar{q}}$ and his three disconneccted configuration energy at temperature $T=0.08$ $\rm{GeV}$. The black line is the energy of $E_{QQ\bar{q}\bar{q}}$, the blue line is the energy of $2E_{Q\bar{q}}$, the orange line is the energy of $E_{QQq}+E_{\bar{q}\bar{q}\bar{q}}$, and the green line is the energy of $E_{Qqq\bar{q}}+E_{Q\bar{q}}$.}
\end{figure}
The results of the calculation are shown in Fig.\ref{fig17}. Here, we define the string-breaking distance as the intersection point of the two energies. As shown in Fig.\ref{fig17}(b), we can see that when $L_{QQ\bar{q}\bar{q}}=0.1870$ $\rm{fm}$, $\rm{QQ\bar{q}\bar{q}}$ will decay into $\rm{Q\bar{q}+Q\bar{q}}$, and when $L_{QQ\bar{q}\bar{q}}=1.3036$ $\rm{fm}$, $\rm{QQ\bar{q}\bar{q}}$ will decay into $\rm{Qq\bar{q}\bar{q}+Q\bar{q}}$.  However, at zero temperature (as shown in Fig.\ref{fig17}(a)), the former decay takes place at $L_{QQ\bar{q}\bar{q}}=0.1874$ $\rm{fm}$, while the latter occurs at $L_{QQ\bar{q}\bar{q}}=1.3147$ $\rm{fm}$.  The first scenario is linked to the process of vertex annihilation, whereas the second pertains to the occurrence of string breaking through the production of light quark pairs. The $\rm{QQ\bar{q}\bar{q}}\longrightarrow \rm{Q\bar{q}+Q\bar{q}}$ is the most possible decay mode. The presence of small temperature will increase the string-breaking distance a little bit.  There is always a energy difference for $\rm{QQ\bar{q}\bar{q}}$ and $\rm{QQq+\bar{q}\bar{q}\bar{q}}$  as shown in Fig.\ref{fig17}.  Thus, the decay mode $\rm{QQ\bar{q}\bar{q}}\longrightarrow \rm{QQq+\bar{q}\bar{q}\bar{q}}$ also will not happen.

\subsection{The relation between \texorpdfstring {$\rm{QQq}$}{} and \texorpdfstring {$\rm{QQ\bar{q}\bar{q}}$}{}}

\begin{figure}
    \centering
    \includegraphics[width=15cm]{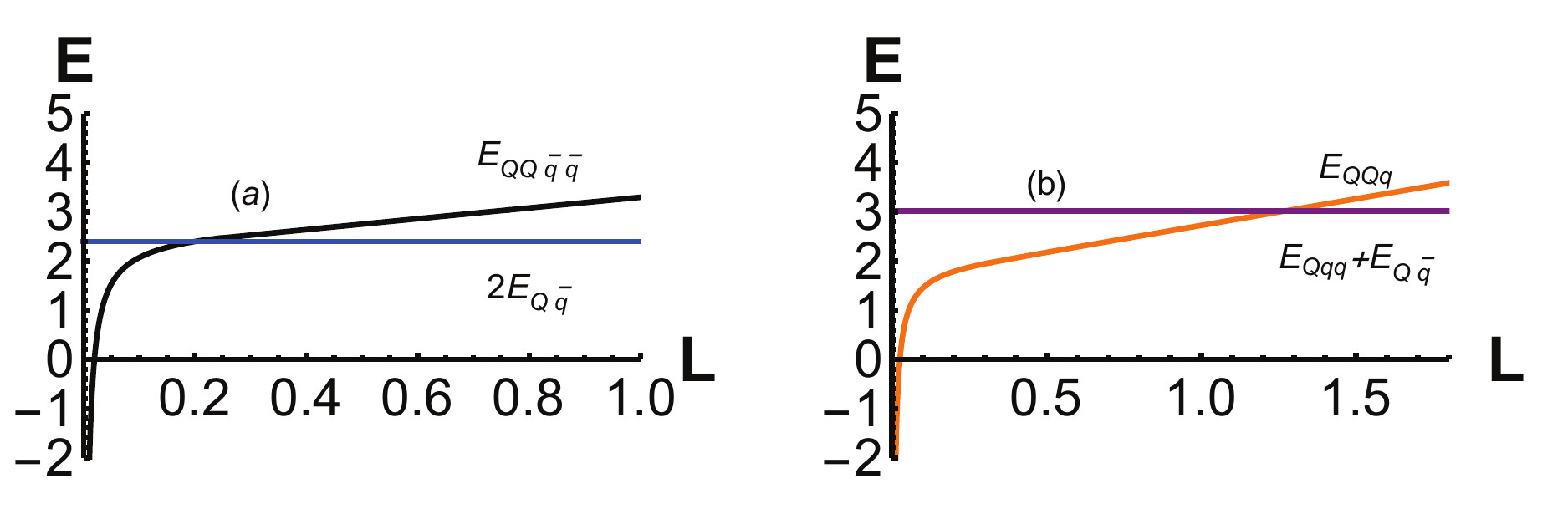}
    \caption{\label{fig18} (a) The energy of the $\rm{QQ\bar{q}\bar{q}}$ and two $\rm Q\bar{q}$. (b) The energy of the $\rm{QQq}$ and  $\rm{Qqq}$+$\rm Q\bar{q}$.}
\end{figure}

In this part, we will discuss the difference between $\rm{QQ\bar{q}\bar{q}}$ and $\rm{QQq}$ at a temperature of $0.08$ $\rm{GeV}$. The $\rm{QQq}$ configuration has also garnered widespread attention. Similar to the $\rm{QQ\bar{q}\bar{q}}$ configuration, the decay process $\rm{QQq\longrightarrow Qqq+Q\bar{q}}$ will occur in the $\rm{QQq}$ configuration at high temperatures.  As shown in Fig.\ref{fig18}, we calculated the two configurations separately as well as the energy after their decay. Then, $\rm{QQ\bar{q}\bar{q}}$ decays to $\rm{Q\bar{q}}$ at $E=2.3806$  $\rm{GeV}$$(L=0.1980$ $\rm{fm})$, and $\rm{QQq}$ decays to $\rm{Qqq+Q\bar{q}}$ at $E=3.0167$  $\rm{GeV}$$(L=1.2646$ $\rm{fm})$. At lower energies and smaller separation distances, the $\rm{QQ\bar{q}\bar{q}}$ configuration will decay, while QQq is more stable. This difference may result from two distinct mechanisms: string breaking by light quarks for $\rm{QQq}$ and string junction annihilation for $\rm{QQ\bar{q}\bar{q}}$.

\begin{figure}
    \centering
    \includegraphics[width=15cm]{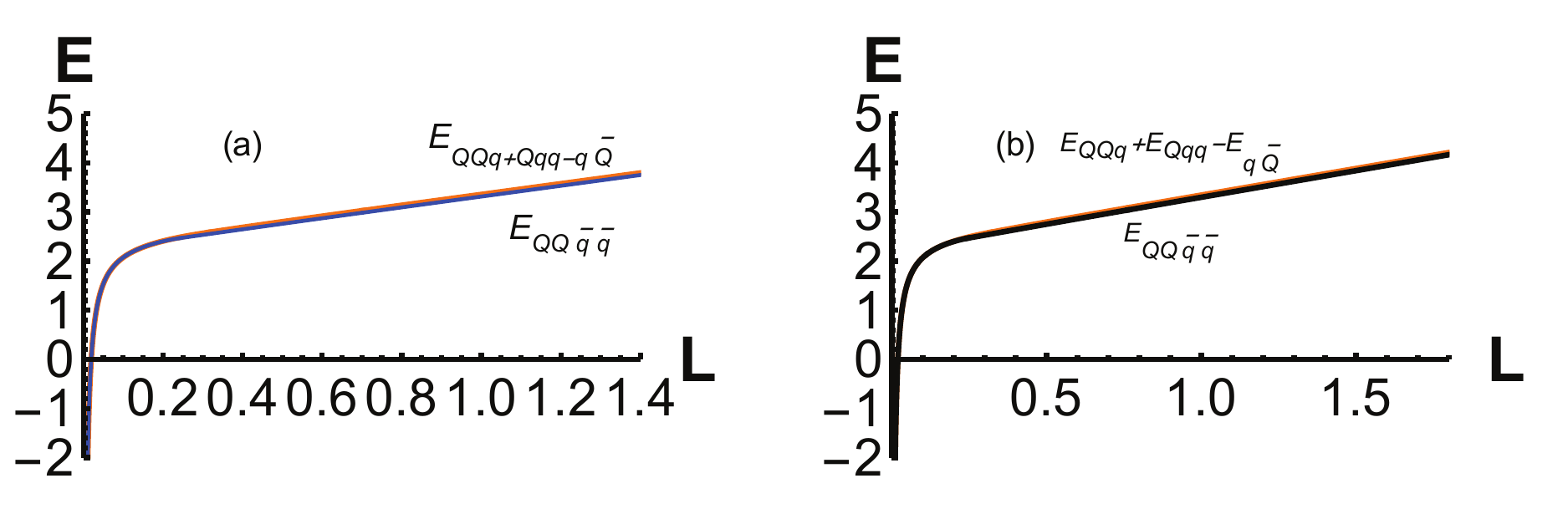}
    \caption{\label{fig19} (a) The energy of $\rm{QQ\bar{q}\bar{q}}$ and $\rm{QQq+Qqq-q{\bar{Q}}}$ at $T=0$ $\rm{GeV}$. (b) The energy of $\rm{QQ\bar{q}\bar{q}}$ and $\rm{QQq+Qqq-q{\bar{Q}}}$ at $T=0.08$ $\rm{GeV}$.}
\end{figure}

Here, we also consider another relation in \cite{13Andreev4kuake,Richard:2018jkw}, which is $E_{\mathrm{QQ \overline {q } \overline { q }}}=E_{\mathrm{QQq}}+E_{\mathrm{Qqq}}-E_{\mathrm{q} \bar{Q}}$. This relationship is derived from heavy quark-diquark symmetry, as illustrated in Fig.\ref{fig19}. Furthermore, similar to the case at zero temperature, we find that relationship occurs at a very small separation distance when $T=0.08$ $\rm{GeV}$. After $L=0.2662$ $\rm{fm}$, the potential energy of $\rm{QQ\bar{q}\bar{q}}$ and $\rm{QQq+Qqq-q\bar{Q}}$ continues to increase. So there's a slight difference in energy, specifically, the energy of $\rm{QQq+Qqq-q\bar{Q}}$ will be slightly higher than that of $\rm{QQ\bar{q}\bar{q}}$. In addition, we find that the energy difference between before and after equation will occur after $L=0.2396$ $\rm{fm}$ when it is at zero temperature, and after $L=0.2662$ $\rm{fm}$ when it is at $T=0.08$ $\rm{GeV}$. The increase in temperature will increase the critical distance for the appearance of energy difference.

\section{summary and conclusion }\label{sec:06}

This paper focuses on the melting and breaking of strings in $\rm{QQ\bar{q}\bar{q}}$ at finite temperatures by using the five-dimensional effective string model, and compares it with QQq under same conditions. During the confinement phase, a dynamic wall which strings cannot penetrate forms below the black-hole horizon.  As the temperature increases, the system enters the deconfinement phase. In this phase, the dynamic wall disappears, causing the $\rm{QQ\bar{q}\bar{q}}$ string to melt and the quarks to become free.  We investigate the energy  of $\rm{QQ\bar{q}\bar{q}}$ at which string melting occurs at four different temperatures.  Subsequently, the study of three decay modes of $\rm{QQ\bar{q}\bar{q}}$ configuration at high temperature is continued, and the temperature at which the different decay modes occur is calculated. Finally, we compare the decay model of the $\rm{QQ\bar{q}\bar{q}}$ configuration at $T=0.08$ $\rm{GeV}$ with that of the QQq and conclude that the QQq configuration is more stable under the same conditions.  Ultimately, we aim to provide additional insights for experiments through our study of the effective string model in the future work.

\section*{Acknowledgments}
This work is supported by the Natural Science Foundation of Hunan Province of China under Grants No. 2022JJ40344, the Research Foundation of Education Bureau of Hunan Province, China (Grant No. 21B0402), and the National Natural Science Foundation of China Grants under No. 12175100.

\section*{References}

\bibliographystyle{unsrt}
\bibliography{ref}

\begin{thebibliography}{10}

\bibitem{1Maldacena:1998TheLargeN}
Juan~Martin Maldacena.
\newblock {The Large N limit of superconformal field theories and
  supergravity}.
\newblock {\em AIP Conf. Proc.}, 484(1):51, 1999.

\bibitem{2Aharony:1999LargeN}
Ofer Aharony, Steven~S. Gubser, Juan~Martin Maldacena, Hirosi Ooguri, and Yaron
  Oz.
\newblock {Large N field theories, string theory and gravity}.
\newblock {\em Phys. Rept.}, 323:183--386, 2000.

\bibitem{3Maldacena:1998}
Juan~Martin Maldacena.
\newblock {Wilson loops in large N field theories}.
\newblock {\em Phys. Rev. Lett.}, 80:4859--4862, 1998.

\bibitem{14Andreev:2019cbc}
Oleg Andreev.
\newblock {Baryon modes in string breaking from gauge/string duality}.
\newblock {\em Phys. Lett. B}, 804:135406, 2020.

\bibitem{4Rey:1998Wilson-Polyakovloopwendu}
Soo-Jong Rey, Stefan Theisen, and Jung-Tay Yee.
\newblock {Wilson-Polyakov loop at finite temperature in large N gauge theory
  and anti-de Sitter supergravity}.
\newblock {\em Nucl. Phys. B}, 527:171--186, 1998.

\bibitem{5Brandhuber:1998limitwendu}
A.~Brandhuber, N.~Itzhaki, J.~Sonnenschein, and S.~Yankielowicz.
\newblock {Wilson loops in the large N limit at finite temperature}.
\newblock {\em Phys. Lett. B}, 434:36--40, 1998.

\bibitem{6Casalderrey-2014wendu}
Jorge Casalderrey-Solana, Hong Liu, David Mateos, Krishna Rajagopal, and
  Urs~Achim Wiedemann.
\newblock {\em {Gauge/String Duality, Hot QCD and Heavy Ion Collisions}}.
\newblock Cambridge University Press, 2014.

\bibitem{Zhang:2015faa}
Zi-qiang Zhang, De-fu Hou, and Gang Chen.
\newblock {Heavy quark potential from deformed $AdS_5$ models}.
\newblock {\em Nucl. Phys. A}, 960:1--10, 2017.

\bibitem{Andreev:2006ct}
Oleg Andreev and Valentine~I. Zakharov.
\newblock {Heavy-quark potentials and AdS/QCD}.
\newblock {\em Phys. Rev. D}, 74:025023, 2006.

\bibitem{Fadafan:2011gm}
K.~Bitaghsir Fadafan.
\newblock {Heavy quarks in the presence of higher derivative corrections from
  AdS/CFT}.
\newblock {\em Eur. Phys. J. C}, 71:1799, 2011.

\bibitem{DeWolfe:2010he}
Oliver DeWolfe, Steven~S. Gubser, and Christopher Rosen.
\newblock {A holographic critical point}.
\newblock {\em Phys. Rev. D}, 83:086005, 2011.

\bibitem{Chen:2021gop}
Xun Chen, Lin Zhang, and Defu Hou.
\newblock {Running coupling constant at finite chemical potential and magnetic
  field from holography *}.
\newblock {\em Chin. Phys. C}, 46(7):073101, 2022.

\bibitem{Zhou:2020ssi}
Jing Zhou, Xun Chen, Yan-Qing Zhao, and Jialun Ping.
\newblock {Thermodynamics of heavy quarkonium in a magnetic field background}.
\newblock {\em Phys. Rev. D}, 102(8):086020, 2020.

\bibitem{Chen:2019rez}
Xun Chen, Danning Li, Defu Hou, and Mei Huang.
\newblock {Quarkyonic phase from quenched dynamical holographic QCD model}.
\newblock {\em JHEP}, 03:073, 2020.

\bibitem{Bohra:2019ebj}
Hardik Bohra, David Dudal, Ali Hajilou, and Subhash Mahapatra.
\newblock {Anisotropic string tensions and inversely magnetic catalyzed
  deconfinement from a dynamical AdS/QCD model}.
\newblock {\em Phys. Lett. B}, 801:135184, 2020.

\bibitem{Alho:2013hsa}
T.~Alho, M.~J\"arvinen, K.~Kajantie, E.~Kiritsis, C.~Rosen, and K.~Tuominen.
\newblock {A holographic model for QCD in the Veneziano limit at finite
  temperature and density}.
\newblock {\em JHEP}, 04:124, 2014.
\newblock [Erratum: JHEP 02, 033 (2015)].

\bibitem{Chen:2020ath}
Xun Chen, Lin Zhang, Danning Li, Defu Hou, and Mei Huang.
\newblock {Gluodynamics and deconfinement phase transition under rotation from
  holography}.
\newblock {\em JHEP}, 07:132, 2021.

\bibitem{Erlich:2005wuwei}
Joshua Erlich, Emanuel Katz, Dam~T. Son, and Mikhail~A. Stephanov.
\newblock {QCD and a holographic model of hadrons}.
\newblock {\em Phys. Rev. Lett.}, 95:261602, 2005.

\bibitem{Chen:2018vty}
Xun Chen, Danning Li, and Mei Huang.
\newblock {Criticality of QCD in a holographic QCD model with critical end
  point}.
\newblock {\em Chin. Phys. C}, 43(2):023105, 2019.

\bibitem{Arefeva:2018hyo}
Irina Aref'eva and Kristina Rannu.
\newblock {Holographic Anisotropic Background with Confinement-Deconfinement
  Phase Transition}.
\newblock {\em JHEP}, 05:206, 2018.

\bibitem{Ewerz:2016zsx}
Carlo Ewerz, Olaf Kaczmarek, and Andreas Samberg.
\newblock {Free Energy of a Heavy Quark-Antiquark Pair in a Thermal Medium from
  AdS/CFT}.
\newblock {\em JHEP}, 03:088, 2018.

\bibitem{Casalderrey-Solana:2011dxg}
Jorge Casalderrey-Solana, Hong Liu, David Mateos, Krishna Rajagopal, and
  Urs~Achim Wiedemann.
\newblock {\em {Gauge/String Duality, Hot QCD and Heavy Ion Collisions}}.
\newblock Cambridge University Press, 2014.

\bibitem{Fang:2015ytf}
Zhen Fang, Song He, and Danning Li.
\newblock {Chiral and Deconfining Phase Transitions from Holographic QCD
  Study}.
\newblock {\em Nucl. Phys. B}, 907:187--207, 2016.

\bibitem{Ding:2015ona}
Heng-Tong Ding, Frithjof Karsch, and Swagato Mukherjee.
\newblock {Thermodynamics of strong-interaction matter from Lattice QCD}.
\newblock {\em Int. J. Mod. Phys. E}, 24(10):1530007, 2015.

\bibitem{Zhou:2021sdy}
Jing Zhou, Xun Chen, Yan-Qing Zhao, and Jialun Ping.
\newblock {Thermodynamics of heavy quarkonium in rotating matter from
  holography}.
\newblock {\em Phys. Rev. D}, 102(12):126029, 2021.

\bibitem{Cai:2012xh}
Rong-Gen Cai, Song He, and Danning Li.
\newblock {A hQCD model and its phase diagram in Einstein-Maxwell-Dilaton
  system}.
\newblock {\em JHEP}, 03:033, 2012.

\bibitem{Li:2011hp}
Danning Li, Song He, Mei Huang, and Qi-Shu Yan.
\newblock {Thermodynamics of deformed AdS$_5$ model with a positive/negative
  quadratic correction in graviton-dilaton system}.
\newblock {\em JHEP}, 09:041, 2011.

\bibitem{Li:2012ay}
Danning Li, Mei Huang, and Qi-Shu Yan.
\newblock {A dynamical soft-wall holographic QCD model for chiral symmetry
  breaking and linear confinement}.
\newblock {\em Eur. Phys. J. C}, 73:2615, 2013.

\bibitem{Richard:2018jkw}
Jean-Marc Richard, Alfredo Valcarce, and Javier Vijande.
\newblock {Doubly-heavy baryons, tetraquarks, and related topics}.
\newblock {\em Bled Workshops Phys.}, 19:24, 2018.

\bibitem{7LHCb:2021vvq}
Roel Aaij et~al.
\newblock {Observation of an exotic narrow doubly charmed tetraquark}.
\newblock {\em Nature Phys.}, 18(7):751--754, 2022.

\bibitem{Drummond:1998ar}
I.~T. Drummond.
\newblock {Strong coupling model for string breaking on the lattice}.
\newblock {\em Phys. Lett. B}, 434:92--98, 1998.

\bibitem{8tetraquark2017}
Jacob Sonnenschein and Dorin Weissman.
\newblock {A tetraquark or not a tetraquark? A holography inspired stringy
  hadron (HISH) perspective}.
\newblock {\em Nucl. Phys. B}, 920:319--344, 2017.

\bibitem{9Andreev:2008some}
Oleg Andreev.
\newblock {Some Multi-Quark Potentials, Pseudo-Potentials and AdS/QCD}.
\newblock {\em Phys. Rev. D}, 78:065007, 2008.

\bibitem{10Najjar:2009da}
Johannes Najjar and Gunnar Bali.
\newblock {Static-static-light baryonic potentials}.
\newblock {\em PoS}, LAT2009:089, 2009.

\bibitem{11Yamamoto:2008jz}
Arata Yamamoto, Hideo Suganuma, and Hideaki Iida.
\newblock {Lattice QCD study of the heavy-heavy-light quark potential}.
\newblock {\em Phys. Rev. D}, 78:014513, 2008.

\bibitem{12Francis:2017bjr}
Anthony Francis, Renwick~J. Hudspith, Randy Lewis, and Kim Maltman.
\newblock {More on heavy tetraquarks in lattice QCD at almost physical pion
  mass}.
\newblock {\em EPJ Web Conf.}, 175:05023, 2018.

\bibitem{13Andreev4kuake}
Oleg Andreev.
\newblock {$\rm QQq\bar{q}$ potential in string models}.
\newblock {\em Phys. Rev. D}, 105(8):086025, 2022.

\bibitem{15Andreev:2015iaa}
Oleg Andreev.
\newblock {Model of the $N$-quark potential in $SU(N)$ gauge theory using
  gauge-string duality}.
\newblock {\em Phys. Lett. B}, 756:6--9, 2016.

\bibitem{Stringbreaking}
John Bulava, Ben H\"orz, Francesco Knechtli, Vanessa Koch, Graham Moir, Colin
  Morningstar, and Mike Peardon.
\newblock {String breaking by light and strange quarks in QCD}.
\newblock {\em Phys. Lett. B}, 793:493--498, 2019.

\bibitem{LiYaodong2023}
Yaodong Li, C.~W. von Keyserlingk, Guanyu Zhu, and Tomas Jochym-O'Connor.
\newblock {Phase diagram of the three-dimensional subsystem toric code}.
\newblock 5 2023.

\bibitem{YiYang}
Yi~Yang and Pei-Hung Yuan.
\newblock {Confinement-deconfinement phase transition for heavy quarks in a
  soft wall holographic QCD model}.
\newblock {\em JHEP}, 12:161, 2015.

\bibitem{Chen:Movingheavyquarkonium}
Xun Chen, Sheng-Qin Feng, Ya-Fei Shi, and Yang Zhong.
\newblock {Moving heavy quarkonium entropy, effective string tension, and the
  QCD phase diagram}.
\newblock {\em Phys. Rev. D}, 97(6):066015, 2018.

\bibitem{Chen:2019Quarkyonicphase}
Xun Chen, Danning Li, Defu Hou, and Mei Huang.
\newblock {Quarkyonic phase from quenched dynamical holographic QCD model}.
\newblock {\em JHEP}, 03:073, 2020.

\bibitem{Zhou:2020Thermodynamicsof}
Jing Zhou, Xun Chen, Yan-Qing Zhao, and Jialun Ping.
\newblock {Thermodynamics of heavy quarkonium in a magnetic field background}.
\newblock {\em Phys. Rev. D}, 102(8):086020, 2020.

\bibitem{Andreev:2006nw}
Oleg Andreev and Valentin~I. Zakharov.
\newblock {On Heavy-Quark Free Energies, Entropies, Polyakov Loop, and
  AdS/QCD}.
\newblock {\em JHEP}, 04:100, 2007.

\bibitem{Andreev:2006TheSpatialString}
Oleg Andreev and Valentine~I. Zakharov.
\newblock {The Spatial String Tension, Thermal Phase Transition, and AdS/QCD}.
\newblock {\em Phys. Lett. B}, 645:437--441, 2007.

\bibitem{HeSong:2010Heavyquarkpotential}
Song He, Mei Huang, and Qi-shu Yan.
\newblock {Heavy quark potential and QCD beta function from a deformed $AdS_5$
  model}.
\newblock {\em Prog. Theor. Phys. Suppl.}, 186:504--509, 2010.

\bibitem{Andreev3kuake}
Oleg Andreev.
\newblock {Some Properties of the $QQq$-Quark Potential in String Models}.
\newblock {\em JHEP}, 05:173, 2021.

\bibitem{Chen3kuake}
Xun Chen, Bo~Yu, Peng-Cheng Chu, and Xiao-hua Li.
\newblock {Studying the potential of QQq at finite temperature in a holographic
  model *}.
\newblock {\em Chin. Phys. C}, 46(7):073102, 2022.

\bibitem{Andreev:2007zv}
Oleg Andreev.
\newblock {Some Thermodynamic Aspects of Pure Glue, Fuzzy Bags and Gauge/String
  Duality}.
\newblock {\em Phys. Rev. D}, 76:087702, 2007.

\bibitem{Witten:1998xy}
Edward Witten.
\newblock {Baryons and branes in anti-de Sitter space}.
\newblock {\em JHEP}, 07:006, 1998.

\bibitem{Witten:1998zhongzidingdian}
Edward Witten.
\newblock {Anti-de Sitter space and holography}.
\newblock {\em Adv. Theor. Math. Phys.}, 2:253--291, 1998.

\bibitem{Aharony:1998fanzhongzidingdian}
Ofer Aharony and Edward Witten.
\newblock {Anti-de Sitter space and the center of the gauge group}.
\newblock {\em JHEP}, 11:018, 1998.

\bibitem{Andreev:2020pqy}
Oleg Andreev.
\newblock {String Breaking, Baryons, Medium, and Gauge/String Duality}.
\newblock {\em Phys. Rev. D}, 101(10):106003, 2020.

\bibitem{Andreev:2016some}
Oleg Andreev.
\newblock {Some Aspects of Three-Quark Potentials}.
\newblock {\em Phys. Rev. D}, 93(10):105014, 2016.

\bibitem{Andreev:2021bfg}
Oleg Andreev.
\newblock {Remarks on static three-quark potentials, string breaking and
  gauge/string duality}.
\newblock {\em Phys. Rev. D}, 104(2):026005, 2021.

\bibitem{Peeters:2006iu}
Kasper Peeters, Jacob Sonnenschein, and Marija Zamaklar.
\newblock {Holographic melting and related properties of mesons in a quark
  gluon plasma}.
\newblock {\em Phys. Rev. D}, 74:106008, 2006.

\bibitem{Hoyos-Badajoz:2006dzi}
Carlos Hoyos-Badajoz, Karl Landsteiner, and Sergio Montero.
\newblock {Holographic meson melting}.
\newblock {\em JHEP}, 04:031, 2007.

\bibitem{Fadafan:2012qy}
K.~Bitaghsir Fadafan and E.~Azimfard.
\newblock {On meson melting in the quark medium}.
\newblock {\em Nucl. Phys. B}, 863:347--360, 2012.

\bibitem{White:2007tu}
C~D White.
\newblock {The Cornell potential from general geometries in AdS / QCD}.
\newblock {\em Phys. Lett. B}, 652:79--85, 2007.

\bibitem{Gribov:1977wm}
V.~N. Gribov.
\newblock {Quantization of Nonabelian Gauge Theories}.
\newblock {\em Nucl. Phys. B}, 139:1, 1978.

\bibitem{Karch:2006pv}
Andreas Karch, Emanuel Katz, Dam~T. Son, and Mikhail~A. Stephanov.
\newblock {Linear confinement and AdS/QCD}.
\newblock {\em Phys. Rev. D}, 74:015005, 2006.

\bibitem{Bazavov:2011nk}
A.~Bazavov et~al.
\newblock {The chiral and deconfinement aspects of the QCD transition}.
\newblock {\em Phys. Rev. D}, 85:054503, 2012.

\end{thebibliography}

\end{document}